\documentclass[fleqn,usenatbib]{mnras}

\usepackage[T1]{fontenc}

\usepackage{graphicx}	
\usepackage{amsmath}	
\usepackage{amssymb}	
\usepackage{makecell}   
\usepackage{xspace}
\usepackage{ulem}

\newcommand{\Msun}{\rm M_{\odot}}

\newcommand{\jDM}{$j_{\rm DM}$\xspace}
\newcommand{\jgas}{$j_{\rm gas}$\xspace}

\newcommand{\xjDM}{j_{\rm DM}}

\newcommand{\xspintot}{\lambda_{\rm tot}}
\newcommand{\xspingas}{\lambda_{\rm gas}}
\newcommand{\xspinDM}{\lambda_{\rm DM}}

\newcommand{\f}{f_g}
\newcommand{\avg}[1]{\langle#1\rangle}

\newcommand{\jg}{j_{\rm gas}}
\newcommand{\jd}{j_{\rm DM}}

\title[Spin transfer in collapsing haloes]{Spin transfer from dark matter to gas during halo formation}

\author[J. Li et al.]{
Jie Li,$^{1}$\thanks{E-mail: jie.li@icrar.org}
Danail Obreschkow,$^{1,2}$
Chris Power,$^{1,2}$ and Claudia del P. Lagos$^{1,2}$ 
\\
\!\!$^{1}$\,International Centre for Radio Astronomy Research, M468, University of Western Australia, 35 Stirling Hwy, Perth, WA 6009, Australia\\
\!\!$^{2}$\,ARC Centre of Excellence for All Sky Astrophysics in 3 Dimensions (ASTRO 3D)}

\date{Accepted XXX. Received YYY; in original form ZZZ}

\pubyear{2021}

\begin{document}
\label{firstpage}
\pagerange{\pageref{firstpage}--\pageref{lastpage}}
\maketitle

\begin{abstract}
In the protogalactic density field, diffuse gas and collision-less cold dark matter (DM) are often assumed sufficiently mixed that both components experience identical tidal torques. However, haloes in cosmological simulations consistently end up with a higher specific angular momentum (sAM) in gas, even in simulations without radiative cooling and galaxy formation physics. We refine this result by analysing the spin distributions of gas and DM in $\sim$50,000 well-resolved haloes in a non-radiative cosmological simulation from the SURFS suite. The sAM of the halo gas on average ends up $\sim$40\% above that of the DM. This can be pinned down to an excess AM in the inner halo ($<$50\% virial radius), paralleled by a more coherent rotation pattern in the gas. We uncover the leading driver for this AM difference through a series of control simulations of a collapsing ellipsoidal top-hat, where gas and DM are initially well mixed. These runs reveal that the pressurised inner gas shells collapse more slowly, causing the DM ellipsoid to spin ahead of the gas ellipsoid. The arising torque generally transfers AM from the DM to the gas. The amount of AM transferred via this mode depends on the initial spin, the initial axes ratios and the collapse factor. These quantities can be combined in a single dimensionless parameter, which robustly predicts the AM transfer of the ellipsoidal collapse. This simplistic model can quantitatively explain the average AM excess of the gas found in the more complex non-radiative cosmological simulation.
\end{abstract}

\begin{keywords}
cosmology: theory -- hydrodynamics  -- physical data and processes
\end{keywords}



\section{Introduction}\label{s:introduction}

Angular momentum (AM) plays a fundamental role in all modern theories of galaxy formation. For example, AM is strongly linked to other galaxy properties such as stellar mass \citep[e.g.][]{fall_formation_1980}, morphology \citep[e.g.][]{romanowsky_angular_2012,cortese_sami_2016}, and gas fraction \citep[e.g.][]{obreschkow_angular_2016}. Thus, understanding how galaxies gain their AM is a central part of understanding galaxy formation. Numerical studies show that many factors can affect the AM of galaxies, including stellar feedback \citep[e.g.][]{defelippis_impact_2017}, gas accretion \citep[e.g.][]{el-badry_gas_2018}, galaxy mergers \citep[e.g.][]{lagos_connection_2018,lagos_quantifying_2018}, along with the primordial spin of the pristine halo gas itself. In fact, in the standard cold dark matter cosmological framework, protogalactic haloes of dark matter (DM) and diffuse gas acquire AM through tidal torques \citep[e.g.][]{peebles_origin_1969}. As the gas in a halo cools and flows to the centre of potential, its AM causes the formation of a spinning gaseous disc \citep[e.g.][]{white_core_1978} in which stars can form via local runaway instabilities. In this simplistic picture, the spin of the diffuse halo gas seeds the AM of the galaxies.

Historically, the starting assumption is that the AM of the halo gas remains conserved as this gas cools and forms a galactic disc. Most semi-analytic models to-date impose some version of conservation \citep[see overview by][]{hou_how_2021}, but high-resolution cosmological simulations cast serious doubts on its validity. It now seems that the gas normally loses a significant amount of AM as it cools onto the disc \citet{stevens_how_2017}, but AM is then increased again via local stellar feedback to form a stable disc \citep[e.g.][]{guedes_forming_2011,zjupa_angular_2017}. Several works have investigated the physical drivers of the increased sAM of cold halo gas and/or star-forming gas \citep[e.g.][]{pichon_rigging_2011,danovich_four_2015,stewart_orbiting_2011,stewart_angular_2013,stewart_high_2017}. For example, \citet{danovich_four_2015} developed a four-step model from their cosmological simulations with sub-grid physics to describe the AM acquisition of galaxies, suggesting that the cold gas in the elongated thin streams outside the halo gains more sAM than the DM, because of the higher quadrupole moment associated with the thin gas streams. As a result, the spin of the cold gas of the galaxy can be 3-times higher than that of DM. Similarly, \citet{stewart_high_2017} compared five high-resolution zoom-in simulations and found that in all simulations the cold filamentary gas accretion onto the halo resulted in about 4-times more sAM in cold halo gas than surrounding DM.

Whereas the AM conservation of the cooling and star-forming gaseous component has come under scrutiny, another important assumption remains almost unchallenged. This is the assumption that, in the protogalactic haloes, DM and gas share the same sAM. This assumption is widely used in analytical and semi-analytic models. However, modern cosmological simulations of gas and DM do not fully support this assumption. They generally find that gas has a higher sAM than DM. \citet{zjupa_angular_2017} found that specific AM (sAM) of the halo gas lies a factor 1.8 above that of the DM in the \textsc{illustris} simulation with full physics turned on. They suggested that this enhancement of the gas spin can be explained by an AM transfer from the DM to the gas during mergers, as well as by stellar feedback removing low AM gas from the haloes.

Interestingly, complex galactic processes, such as radiative cooling and star formation, are not the only culprit behind the difference in the sAM of halo DM and gas. Even cosmological simulations of DM and an {\it ideal} gas, without radiative cooling and galaxy formation physics \citep{chen_angular_2003,sharma_angular_2005,zjupa_angular_2017}, consistently showed that the ratio between the mean spin parameters of gas and DM, which is equal to the ratio of the mean sAM, lies around 1.3--1.4 at redshift $z=0$. This ratio gradually increases with cosmic time. For instance, \citet{van_den_bosch_angular_2002} showed in their own non-radiative hydrodynamical simulation that the spin parameter distributions of gas and DM are nearly identical at $z=3$. Similarly, \citet{sharma_angular_2005} showed, based on 41 simulated haloes, that the spin ratio increases from $\sim1.1$ at $z=4$ to $1.4$ at $z=0$. The analysis of \citet{zjupa_angular_2017} corroborates this qualitative trend and expands on it by showing that the median spin parameter of DM remains nearly constant with time, while the median gas spin gradually increases.

To explain the enhancement of gas spin seen in non-radiative simulations, \citet{sharma_origin_2012} suggested that DM has an inside-out transfer of AM, due to dynamical friction, while gas has an outside-in transfer of AM, as the low AM gas in the inner parts shocks with high AM gas falling in from the outer parts. Then the outer part of the gas moves out of the virial radius as time progresses. \citet{zjupa_angular_2017} argued that in such a scenario, the gas-to-DM spin ratio would be expected to be similar when the outer regions are fully included, such as in the case of friends-of-friends (FOF) haloes. However, they found the FOF haloes also have an enhanced gas-to-DM spin, suggesting that additional mechanisms need to be invoked in the explanation. They comment that gas continuously acquires sAM throughout cosmic time. This acquisition could be explained by mergers getting ram pressure stripped during halo infall, inducing a decoupling between gas and DM and producing torques between these two components as a consequence. This allows a net transfer of angular momentum from dark matter to gas. But this scenario cannot explain the direction of AM transformation: why is AM preferentially  transfer from DM to gas, rather than the other way around?

This paper sets out to identify the main cause behind the average AM excess in gas ($\avg{\jg}/\avg{\jd}>1$) in cosmological simulations of structure formation {\it without} radiative gas cooling and galaxy formation. We found that a combination of such a simulation with a set of control simulations of collapsing top-hat ellipsoids holds the clues to an explanation that seems both intuitive and numerically dominant. The control simulations reveal a significant AM transfer from DM to gas in the inner halo during the collapse phase, which appears to be a universal mechanism also seen in realistic haloes in cosmological simulations.

This paper is structured as follows. We begin in Section~\ref{s:cosmosims} with a description of our cosmological simulation and the post-processing. In Section~\ref{s:controlsims}, we introduce a set of control simulations of an ellipsoidal top-hat collapse, which provide a detailed view of how AM transfers from DM to gas in this simplified setting. Section~\ref{s:discussion} links the results of the control simulations to the cosmological simulation, which involves the introduction of a dimensionless formulation of spin transfer. We summarise our key findings in Section~\ref{s:conclusions}.

\section{Spin in Cosmological Simulations}
\label{s:cosmosims}

\subsection{Non-radiative cosmological simulation}

To study large-scale AM differences between DM and gas in a $\Lambda$CDM cosmological context, we rely on a simulation from the Synthetic UniveRses For Surveys (SURFS) simulation suite \citep{elahi_surfs_2018}, run with the massively parallel simulation code \textsc{Gadget}-2 \citep{springel_cosmological_2005}. The relevant cosmological parameters are $\Omega_{\rm M}=0.3121$, $\Omega_{\rm b}=0.0489$, $\Omega_{\Lambda}=0.6879$ and a dimensionless Hubble parameter $h=0.6751$, defined such that the Hubble constant is $H_0=100 h\rm\,km\,s^{-1}\,Mpc^{-1}$.

The SURFS run used in this paper (L210N1024NR) is a cosmological gravitational+hydrodynamics $N$-body simulation of a cubic volume with a comoving side length of 210$\,h^{-1}\,$Mpc. It uses $2\times1024^3$ particles, half of which are collision-less DM particles of mass $6.29\times 10^8\,h^{-1}\Msun$ and half are gas particles of mass $1.17\times 10^{8}\,h^{-1}\Msun$. The gas is subject to an ideal equation of state of adiabatic index 5/3 (monoatomic). This gas can only change its internal energy adiabatically via work by gravitational and pressure forces, but it cannot cool radiatively, which suppresses the formation of discs. The simulation outputs are stored in 200 snapshots, equally spaced in logarithmic expansion factor by 0.007~dex.

\subsection{Halo sample}
\label{ss:sample}

In each simulation snapshot, haloes are identified using the \textsc{VELOCIraptor} code \citep{elahi_hunting_2019}, which is a massively parallel galaxy (sub)halo finder based on 3D/6D FOF algorithm \citep{davis_evolution_1985}. \textsc{VELOCIraptor} follows a two-step approach: (1) identify FOF groups and (2) identify substructure within each FOF group, optionally using phase-space information.

In this work we use the first-generation haloes, here defined as FOF groups, but we remove self-bound substructure. This choice to remove substructure was motivated by the fact that a non-negligible fraction FOF groups look like a collection of loosely connected, well-separated structures, whose common centre of mass is far from either centre of potential, e.g.~dumbbell-like structures. This can be a disturbing feature in studying the spin of individual haloes, due to the significance of orbital AM between these loosely connected parts. Removing substructure leaves us with more coherent haloes, whose centre of mass generally lies close to the centre of potential. Of course, removing substructure also automatically removes smaller self-bound subhaloes that orbit inside the virial zone of a parent halo. However, through a series of tests with different \textsc{VELOCIraptor} settings (3D vs 6D, with and without substructure), we found that different choices of identifying haloes and handling their subhalo population only changes the results of this paper insignificantly.

Along with identifying the haloes, \textsc{VELOCIraptor} returns various halo properties, as described in Appendix~B of \citet{elahi_hunting_2019}. In the following sections, we mainly use the total halo mass $M_{\rm halo}$, defined as the FOF mass without substructure, the virial mass $M_{\rm vir}$ and the virial radius $R_{\rm vir}$, defined as the radius of a spherical region with a mean density contrast of 200.

For our analysis of AM, we require haloes that are kinematically well-resolved both in terms of their DM and gas components. We therefore restrict our analysis to haloes with a mass of $M_{\rm halo}\geq h^{-1}\times10^{12}\Msun$, corresponding to $\gtrsim1,000$ particles of each type (DM and gas). Following the quantitative tests of \citet{obreschkow_characterizing_2020,correa_accretion_2015}, this number of particles allows for halo assembly histories to be reasonably well resolved for most dynamical properties to be converged. The number of haloes resulting from this selection is 47,972 ($z=0$), 38,124 ($z=1$), 21,168 ($z=2$), 8,578 ($z=3$) and 2,298 ($z=4$).

\subsection{Global spin distributions}
\label{ss:GlobalAM}

The AM of haloes is conveniently expressed in terms of the scale-free dimensionless spin parameter \citep{peebles_origin_1969},
\begin{equation}
    \lambda=\frac{J|E|^{1/2}}{GM^{5/2}},
    \label{equ:lambda_P}
\end{equation}
where $J$ is the total AM (vector norm) relative to the centre of mass, $E=E_{\rm kin}+E_{\rm pot}$ is the total mechanical energy, $G$ is the gravitational constant and $M$ is the total mass. However, calculating the potential energy $E_{\rm pot}$ is computationally expensive for a large number of particles. This difficulty is bypassed by the alternative spin parameter \citep{bullock_universal_2001},
\begin{equation}
    \lambda'=\frac{j}{\sqrt{2}R_{\rm vir} V_{\rm vir}},
    \label{equ:lambda_B}
\end{equation}
where $R_{\rm vir}$ and $V_{\rm vir}$ are the virial radius and the circular velocity at the virial radius, and $j=J/M_{\rm halo}$ is the sAM of the halo. The factor $1/\sqrt{2}$ in $\lambda'$ is introduced such that $\lambda'=\lambda$ in the case of a singular isothermal sphere truncated at $R_{\rm vir}$. In more realistic haloes, $\lambda'$ tends to be a bit lower than $\lambda$. For example, \citet{bullock_universal_2001} found a median values of $\lambda'\approx 0.035$ and $\lambda\approx 0.042$ in their non-radiative cosmological simulation. This difference should be taken into account in precision measurements, however, it can be mostly ignored in this paper, because the spin ratio between gas and DM barely depends on the choice of the spin parameter.\

We will use the Bullock spin parameter $\lambda'$ in the analysis of the cosmological simulation, where virial radii and masses are well defined. On the other hand, we will use the Peebles spin parameter $\lambda$ in our control simulations described in Section~\ref{s:controlsims}, since the virial quantities are not well defined in these non-cosmological (and non-expanding) simulations.

We compute this parameter separately for DM ($\xspinDM'$), gas ($\xspingas'$) and both components combined ($\xspintot'$), by substituting $j$ in equation~\ref{equ:lambda_B} for $\xjDM=J_{\rm DM}/M_{\rm DM}$ and so on, but keeping $R_{\rm vir}$ and $V_{\rm vir}$ defined for the total halo mass in SURFS. Note that we use the vector norm of the individual AM, irrespective of their alignment.

Table~\ref{tab:spin} shows the mean values of $\xspintot'$, $\xspingas'$ and $\xspinDM'$, as well as the ratio of $\avg{\xspingas'}/\avg{\xspinDM'}$ from $z=4$ to $z=0$. We deliberately use $\avg{\xspingas'}/\avg{\xspinDM'}$ rather than $\avg{\xspingas'/\xspinDM'}$ to avoid convergence issues caused by some values $\xspinDM'\approx0$. Incidentally, such values close to zero cause both $\avg{\xspingas'/\xspinDM'}$ and $\avg{\xspinDM'/\xspingas'}$ to be larger than unity, making their interpretation cumbersome.

Interestingly, $\avg{\xspingas'}/\avg{\xspinDM'}$ increases through cosmic time and reaches 1.44 at $z=0$ in SURFS. This value is close to the $\avg{\xspingas'/\xspinDM'}$ of 41 haloes in \citet{sharma_angular_2005} Figure~3. They obtained $\avg{\xspingas'/\xspinDM'}_{z=4}=1.10$ and $\avg{\xspingas'/\xspinDM'}_{z=0}=1.42$ from their halo sample. \citet{zjupa_angular_2017} also found a similar physical trend in $\sim65000$ FOF haloes in \textsc{illustris-2-NR}, where the ratio of the median value of the Peebles spin parameter is 1.308 at $z=0$ and 0.97 at $z=8$, confirming that gas and DM have identical `initial' spin. \citet{gottlober_shape_2007} report a Bullock spin parameter ratio of $\avg{\lambda'_{\rm{gas},z=0}/\lambda'_{\rm{DM},z=0}}=1.39$  from more than 10000 cluster sized FOF haloes with masses larger than $5\times 10^{13}\,h^{-1}\Msun$.

\begin{figure}
    \centering
    \includegraphics[width=\columnwidth]{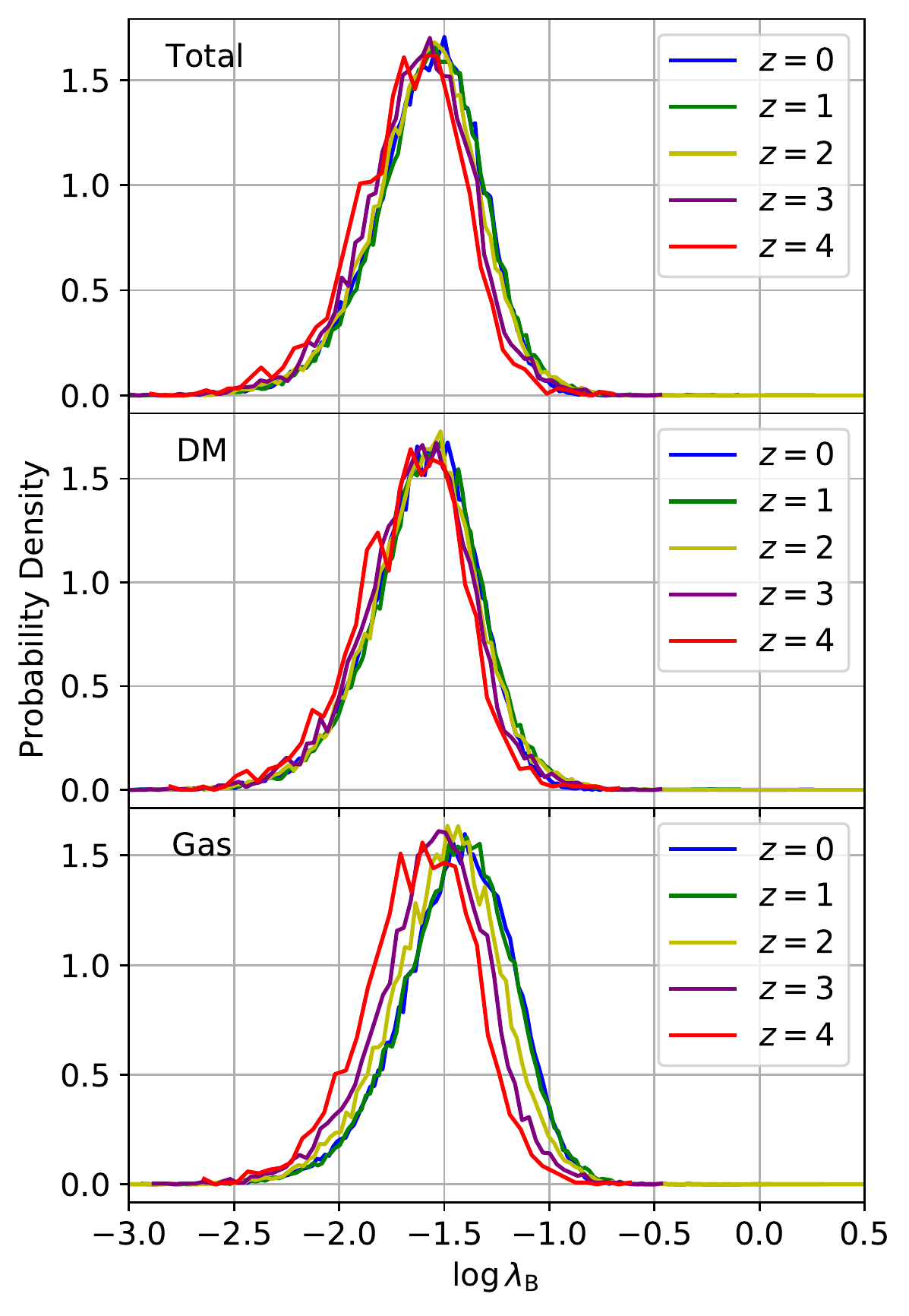}
    \caption{\textit{Top panel:} Spin parameter distributions of all materials in haloes at different redshifts. \textit{Middle panel:} Spin parameter distributions of DM in haloes at the same redshifts. \textit{Bottom panel:} Spin parameter distributions of gas in haloes at the same redshifts.}
    \label{fig:spin_evo}
\end{figure}

The evolution of the spin parameter distributions from $z=4$ to $z=0$ of $\xspintot'$, $\xspinDM'$ and $\xspingas'$ is shown in Figure~\ref{fig:spin_evo}. The top panel shows that the spin parameter distribution is nearly log-normal and remains almost constant with time, as found in previous studies \citep[e.g.][]{knebe_correlation_2008}. The spin parameter distribution of the DM also remains approximately constant (middle panel). On the contrary, the gas spin parameter distribution shifts to higher values with cosmic time. As a consequence, the ratio $\avg{\xspingas'}/\avg{\xspinDM'}$ increases over time, as can also be seen in Table~\ref{tab:spin}.

\begin{figure}
    \centering
    \includegraphics[width=\columnwidth]{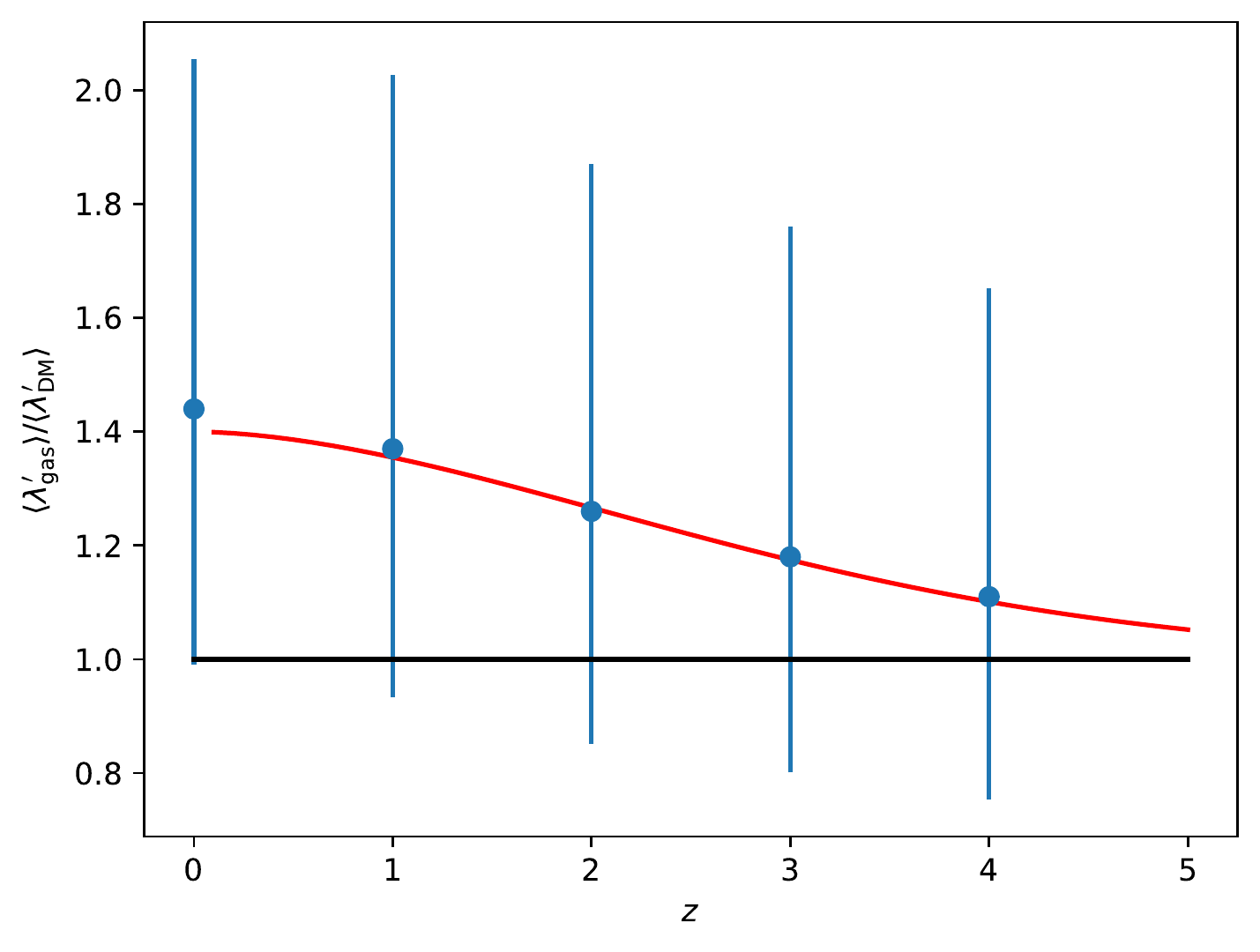}
    \caption{$z-\avg{\xspingas'}/\avg{\xspinDM'}$ relation from SURFS. The blue points present the ratio of $\avg{\xspingas'}/\avg{\xspinDM'}$ at different redshifts. Vertical bars represent the 16th to 84th quantile ranges of the individual $\xspingas'/\xspinDM'$ values at each redshift. The red curve shows the fit of equation (\ref{equ:jratio_z_fit}), which is asymptotes to unity (black line) at a high $z$.}
    \label{fig:lambda_redshift}
\end{figure}

The numerical redshift dependence of $\avg{\xspingas'}/\avg{\xspinDM'}$ is well approximated by the heuristic equation
\begin{equation}
    \avg{\xspingas'}/\avg{\xspinDM'} = 1+ae^{-bz^{c}},
    \label{equ:jratio_z_fit}
\end{equation}
with $a=0.4\pm 0.002$, $b=0.12\pm 0.003$, $c=1.76\pm 0.03$ being the maximum likelihood solutions and their respective 1-$\sigma$ uncertainties in the Laplace approximation. This fit is shown in Figure~\ref{fig:lambda_redshift}. This relation can be used to help SAMs to better estimate $\xspingas$ in DM-only simulations, hence improving on the standard assumption that the halo gas has the same spin as the DM.

\begin{table}
    \centering
    \begin{tabular}{cccccc}
       \hline
       $z$  & $\langle \lambda'_{\rm tot} \rangle$ & $\langle \lambda'_{\rm gas} \rangle$ & $\langle \lambda'_{\rm DM} \rangle$ & $\langle \lambda'_{\rm gas} \rangle/\langle \lambda'_{\rm DM} \rangle$ \\
    
       \hline
       0 & 0.0305 & 0.0424 & 0.0293 & 1.44 \\
       1 & 0.0318 & 0.0425 & 0.0309 & 1.37 \\
       2 & 0.0305 & 0.0380 & 0.0301 & 1.26 \\
       3 & 0.0280 & 0.0331 & 0.0280 & 1.18  \\
       4 & 0.0253 & 0.0285 & 0.0255 & 1.11 \\
       \hline 
    \end{tabular}
    \caption{Spin parameter evolution of haloes in the SURFS simulation L210N1024NR. Column 1: the redshifts; Column 2: the mean value of spin parameter of haloes; Column 3: the mean value of spin parameter of gas; Column 4: the mean value of spin parameter of DM; Column 5: the ratio between mean spin parameter of gas to that of DM.}
    \label{tab:spin}
\end{table}

\begin{figure}
    \centering
    \includegraphics[width=\columnwidth]{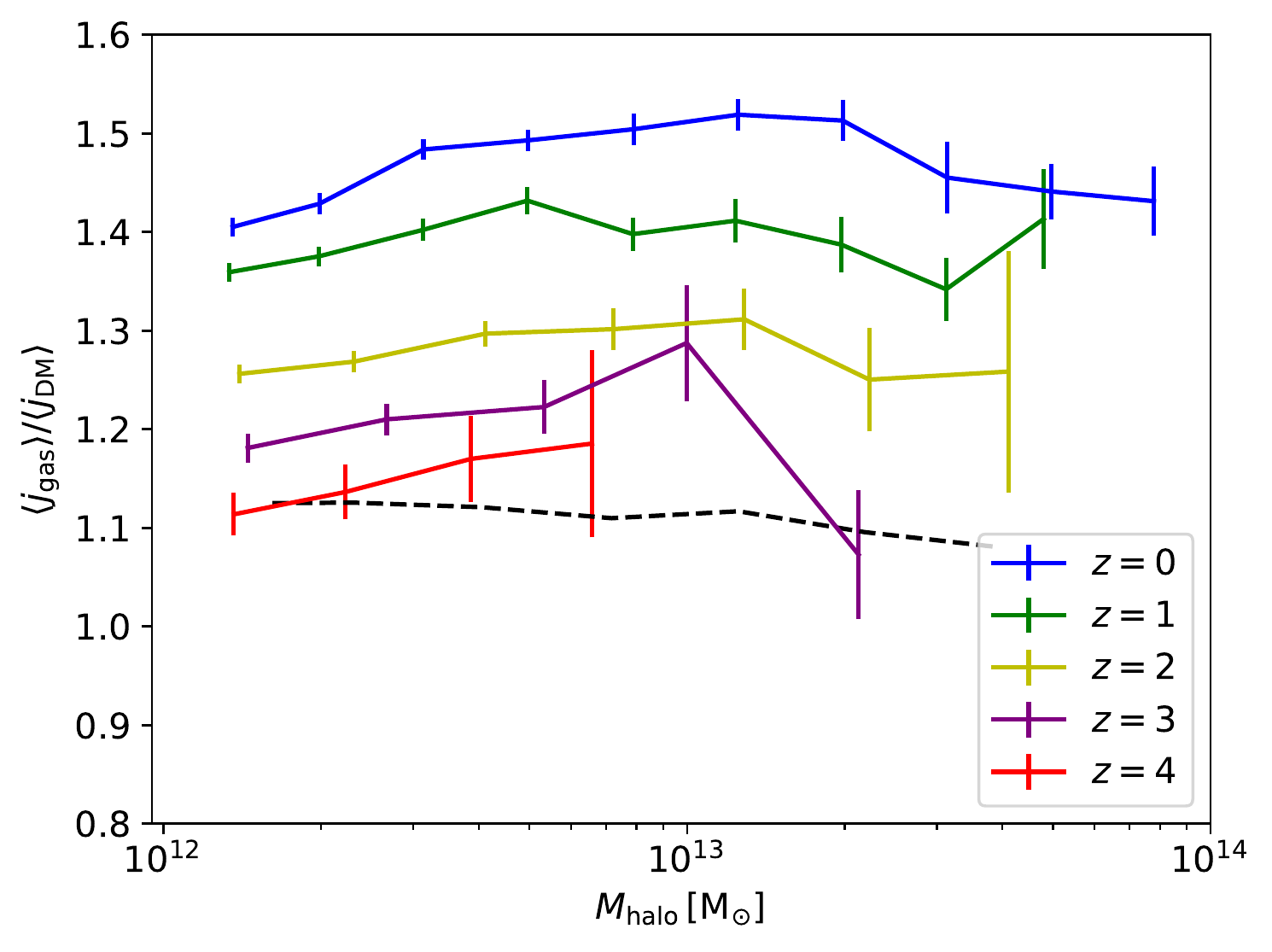}
    \caption{Halo mass-\jgas/\jDM relation at different redshifts. The errorbars present the jacknife uncertainty of the mean value of each bin. Black dashed line shows the \jgas/\jDM at the turnaround point $z_{\rm turn}$, which is discussed in Section~\ref{s:discussion}.}
    \label{fig:SURFSjratio_evo}
\end{figure}

To test if $\xspingas'/\xspinDM'\equiv\jg/\jd$ is mass-dependent, we show $\avg{\jg}/\avg{\jd}$ as a function of $M_{\rm halo}$ at different redshifts in Figure~\ref{fig:SURFSjratio_evo}. At fixed redshift, $\avg{\jg}/\avg{\jd}$ exhibits only a small mass-dependence as expected from the nearly scale-free physics, whose only mass-dependence comes from the relation between mass, formation redshift and concentration \citep{ludlow_mass-concentration-redshift_2016}. The small fluctuations between adjacent mass-bins in Figure~\ref{fig:SURFSjratio_evo} are explainable by statistical uncertainties, represented by the error bars. Not shown in Figure~\ref{fig:SURFSjratio_evo}, but worth emphasising, is the fact that individual values of $\jg/\jd$ vary immensely between different haloes, typically spanning a range from about 0.5 to 5, even at fixed mass. This large halo-to-halo variation comes from the different environments and assembly histories. An expanded discussion of this spread in $\jg/\jd$ will be part of a subsequent paper (see also Section~\ref{s:discussion}).

The primary takeaway point from Figure~\ref{fig:SURFSjratio_evo} is that the ratio $\avg{\jg}/\avg{\jd}$ increases with decreasing redshift. This qualitatively and quantitatively agrees with the values quoted in earlier studies (see Section~\ref{s:introduction}).

Interestingly, most (70\%--80\%) of this AM enhancement in the gas takes place {\it after the turn-around time}, when the root mean square (rms) of the physical radius of the matter ending up in the final halo reaches its maximum. This moment, when the haloes decouple from the cosmic expansion and start collapsing depends on the halo mass and lies mostly between $z\sim1$ and $z\sim3$ in our sample. The ratio $\avg{\jg}/\avg{\jd}$ at this (halo-dependent) turn-around point is plotted as dashed line in Figure~\ref{fig:SURFSjratio_evo}. It is close to 1.1 irrespective of halo mass. This finding is a hint that the AM enrichment of the gas is dominated by processes related to the halo collapse and/or secondary growth, rather than by primordial large scale tidal torques that build up most of the overall AM \citep{peebles_origin_1969,doroshkevich_spatial_1970,barnes_angular_1987}.

\subsection{AM distribution within haloes}
\label{ss:AMdistribution}

In order to find further clues for the origin of the AM enhancement of the halo gas, we  will now look at the radial distribution of AM inside haloes. To this end we use the selected 47,972 SURFS haloes at $z=0$ and bin the particles of each halo into 20 radial shells between the halo centre and the virial radius. We then compute average mass and AM properties in each shell, for gas, DM and both components together.

\begin{figure}
    \centering
    \includegraphics[width=\columnwidth]{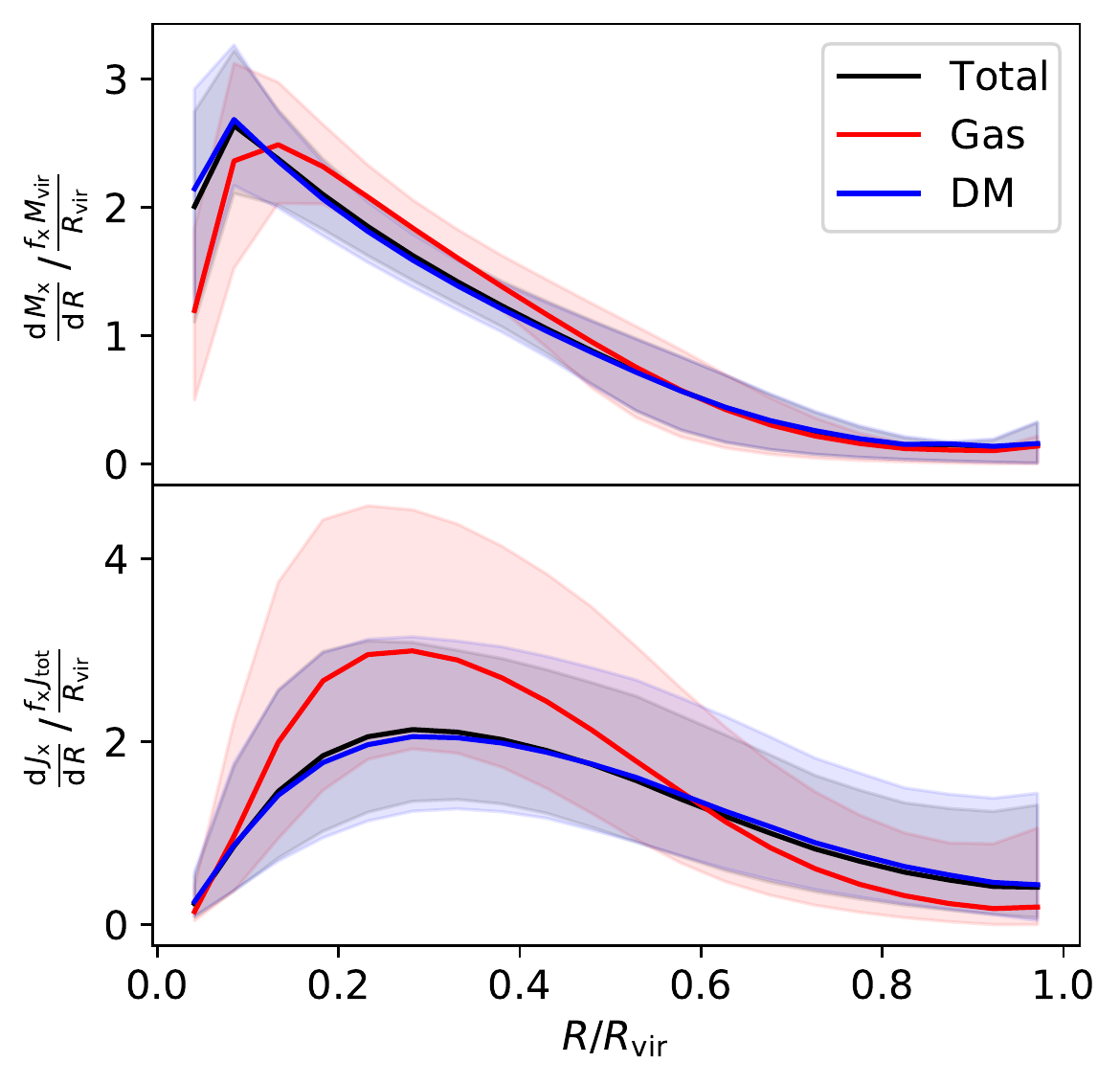}
    \caption{\textit{Top panel}: Mass distribution of each component. The mass distributions of gas and DM are similar. Most of the mass concentrates on the inner halo.  \textit{Bottom panel}: AM distribution of each component. Gas has excess AM comparing with DM at $R<0.6R_{\rm vir}$, larger than the excess AM of DM comparing with gas at $R>0.6R_{\rm vir}$.}
    \label{fig:jmean_M_bin}
\end{figure}

Figure~\ref{fig:jmean_M_bin} shows the average normalised mass and AM distribution within the haloes as a function of radius, normalised by the virial values. The normalised mass in each shell is calculated as
 \begin{equation}
     \frac{{\rm d}M_{\rm x}}{{\rm d} R}\,/\,\frac{f_{\rm x}\,M_{\rm vir}}{R_{\rm vir}}.
 \end{equation}
Here, $x$ represents either the gas, DM or both. $M_x$ and $f_x$ are the mass and mass faction of each component, respectively. $M_{\rm vir}$ is the virial mass of the halo. (Thus, for both components, $M_x=M_{\rm vir}$ and $f_x=1$). Likewise, the normalised AM in each shell is calculated as
\begin{equation}
    \frac{{\rm d}J_{\rm x}}{{\rm d}R}\,/\,\frac{f_{\rm x}\,J_{\rm vir}}{R_{\rm vir}},
\end{equation}
where $J_{\rm x}$ is the total AM of the respective component and $J_{\rm vir}$ is the total AM inside the virial sphere.

As expected, Figure~\ref{fig:jmean_M_bin} shows that the shell with the highest mass contribution lies around $0.1R_{\rm vir}$, i.e.\ near the characteristic concentration radius. The gas is pushed to slightly larger radii by it hydrostatic pressure. In turn, the shell contributing most of the AM sits at a larger radius, near $0.3R_{\rm vir}$, due to the radial term in the definition of AM.

Figure~\ref{fig:jmean_M_bin} clearly shows the excess in $J_{\rm gas}$ over $J_{\rm DM}$, discussed at the halo level in the previous subsections. Interestingly, the radial analysis reveals that most of this excess comes from the inner halo regions at $\lesssim0.5R_{\rm vir}$. This excess of gas AM is far larger than the excess of DM AM at $>0.6R_{\rm vir}$, thus the AM of gas becomes greater than that of DM in the whole halo. Note that previous studies have analysed the sAM profile of DM haloes, e.g. \citealp[][]{bullock_universal_2001,liao_universal_2017} revealing that the outsides of the haloes ($R>0.5R_{\rm vir}$) have higher sAM than the inner parts ($R<0.5R_{\rm vir}$). We also find a similar result using SURFS, however, we show the $J$-profile (rather than the $j$-profile), since only $J$ is an extensive quantity that can be added up.

The excess of $J_{\rm gas}$ relative to $J_{\rm DM}$ at $R=0.1R_{\rm vir}$--$0.5R_{\rm vir}$ is far larger than the excess of $M_{\rm gas}$ relative to $M_{\rm DM}$ at these radii. This implies that even the sAM of the gas lies well above that of the DM. How is this possible? How can the gas hold more sAM than DM at an identical radius? This question arises because one would naturally expect that the velocity of both components is set by the same potential, but the same radius and the same velocity would imply the same sAM. While it is true that the sAM per gas particle is therefore expected to be equal to the sAM per DM particle at the same radius, the mean sAM in a whole shell depends on how well the individual orbits are aligned.

\begin{figure}
    \centering
    \includegraphics[width=\columnwidth]{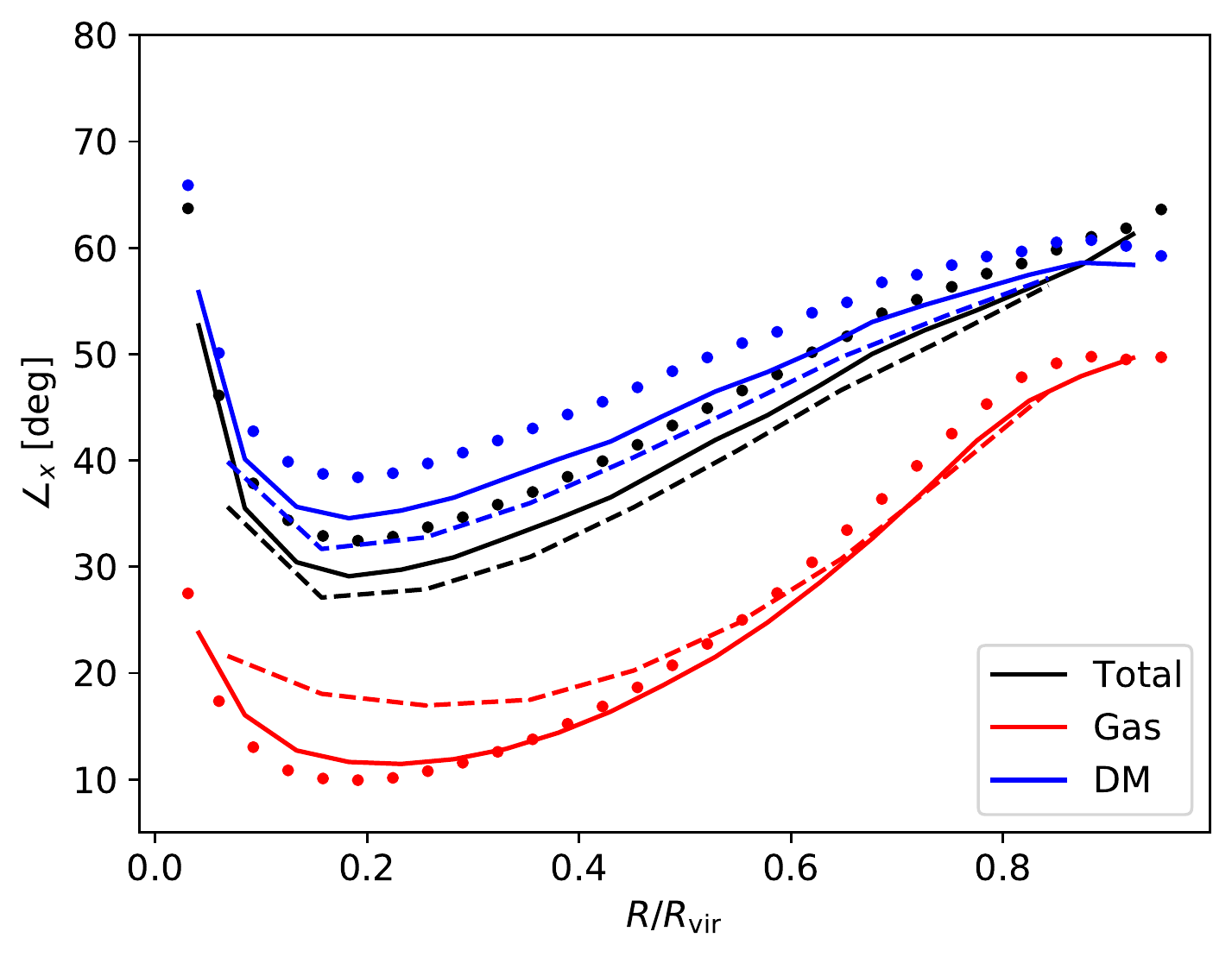}
    \caption{Angle between the spin vectors in adjacent spherical shells. The larger of the angle, the more misaligned a component is with itself. Black, red, blue represent total components, gas and DM respectively. Dashed, solid and dotted lines represent the haloes binned into 10, 20, and 30 bins, respectively.}
    \label{fig:Misalign_shell_shell}
\end{figure}

It turns out that the excess of AM in gas relative to DM is largely explained by the more coherent flow of the gas. In other words, the orbits of different gas particles are more aligned than those of individual DM particles, such that the individual AM vectors of gas particles add up to a larger total vector than for the DM. Figure~\ref{fig:Misalign_shell_shell} shows the angle $\angle_x$ between the AM vector of component $x$ (=gas, DM or both) of adjacent spherical shells of thickness $\Delta R=R_{\rm vir}/20$ (solid lines), averaged over all halos at $z=0$ in our sample. Clearly, the DM exhibits a much more significant internal misalignment than the gas and this conclusion does not depend on the thickness of the shells (see dashed and dotted lines). One might have expected this from the dynamic pressure, $\tfrac{1}{2}\rho\Delta v^2$, that a gas region of density $\rho$ exerts onto another region of relative speed $\Delta v$. Thus, even a frictionless ideal gas has a tendency to produce coherent flows. 

It is also very interesting, but not particularly relevant in our context, to study the alignment between gas and DM. Readers interested in this topic may find useful details in the numerical studies of for instance, non-radiative simulation from \citet{van_den_bosch_angular_2002,sharma_origin_2012,bett_angular_2010}, who found that the total AM of the two components tends to be misaligned between $20^\circ$ and $30^\circ$.

We conclude that the global excess of spin in gas relative to DM (about a factor 1.4 at $z=0$) can be pinned down to an excess of sAM in the gas in the {\it inner} ($<0.5R_{\rm vir}$) region of the haloes, which is possible by virtue of more coherent rotation structures in the gas. However, this explanation only replaces the global phenomenology by a radially resolved picture. It does {\it not} explain, {\it how} the gas acquires more sAM than the DM in the first place. For this to be the case, there has to be a transfer of AM from DM to gas within haloes, or the gas must be able to acquire more AM than DM via external torques.

\section{Control simulations}
\label{s:controlsims}

The findings that, in non-radiative cosmological simulations, most of the excess of AM in gas over DM occurs after the turn-around time of the halo (section~\ref{ss:GlobalAM}) and tends to be localised in the inner halo regions (section~\ref{ss:AMdistribution}) both hint at AM being transferred internally from the DM to the gas inside haloes, during their collapse and/or virialisation phase. However, shining more light on this idea using cosmological simulations is difficult, because it is not easy to determine where the AM of a simulation particle comes from and the complex geometry of real haloes can obscure the leading physical mechanisms. In this section, we therefore turn towards highly idealised control simulations of the formation of single haloes.

One of the simplest models of the formation of a single halo is the spherical top-hat collapse model, in which a sphere of uniform density, initially at rest, is let to collapse \citep[see details in Chapter 6 of][]{mo_galaxy_2010}. While this model has historically led to a lot of first-order insight (e.g. the number density of virialised haloes, the velocity distribuitions), its spherical symmetry does not allow to transfer AM between DM and gas. Even if stochastic density fluctuations were included, it is hard to see how they would tend to asymmetrically enhance the AM of one component. We therefore resort to one of the simplest globally anisotropic models, the model of an ellipsoidal top-hat over-density.

\begin{figure}
    \centering
    \includegraphics[width=\columnwidth]{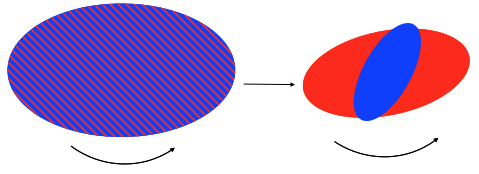}
    \caption{Schematic mechanism to explain how DM transfers AM to gas during an ellipsoidal top-hat collapse. This proto-halo rotates in the anticlockwise direction. Initially, the DM (blue) and gas (red) are well mixed, but the pressure forces slow down the collapse of the gas relative to the DM. By conservation of AM, the DM therefore spins up faster and produces a quadrupole field that torques the gas and transfers AM to it.}
    \label{fig:schematic}
\end{figure}

The basic idea of how an ellipsoidal collapse can transfer AM from the DM to the gas is illustrated in Figure~\ref{fig:schematic}. The rotating ellipsoid starts collapsing due to gravity. Initially, the DM and gas remain congruent, but as the gas density gradually increases, the gas pressure raises (as $P\propto\rho^{\gamma}$) and slows down the collapse of the gas relative to the DM. This leads to a difference in their inertial tensors, such that the mean angular velocity of the DM grows faster than that of the gas. Thus, the DM ellipsoid starts to spin ahead of the gas ellipsoid (Figure~\ref{fig:schematic}, right). The torque thus generated, generally transfers AM from the DM to the gas, unless the angle between their major axes grows above 90 degrees. As the halo virialises through multiple shell-crossings, its overall geometry becomes increasingly spherical, so that the difference in AM between the two phases becomes frozen (up to inefficient small-scale torques at the particle level).

In the following, idealised simulations of an ellipsoidal top-hat collapse with DM and gas will help us quantify and further elaborate on this simple idea of AM transfer.

\subsection{Setting up Control Simluations}
\label{ss:ctl_setup}

Our control simulations of halo formation from an ellipsoidal top-hat collapse were run using the \textsc{Gadget-3} code \citep[a derivative of \textsc{Gadget-2} described in][]{springel_cosmological_2005}. Each individual simulation starts from a single isolated ellipsoid of mass $M$, sampled uniformly with $10^5$ particles of identical mass. A random sub-set of 83 percent of these particles are DM particles, while the remaining 17 percent are gas particles. This sampling resolution suffices for the ratio $j_{\rm gas}/j_{\rm DM}$ to be well modelled and subject only to a small statistical scatter, as demonstrated by the convergence analysis in Appendix \ref{app:convergence}. To keep track of the magnitude of the statistical scatter, we have run five random realisations of each control run presented in the following.

The gas is modelled as an ideal gas of adiabatic index $\gamma=5/3$ without additional gas physics, such as radiative cooling and phase transitions, like in the cosmological simulation discussed in Section~\ref{s:cosmosims}. The gas is initially at a low temperature of 10 K.

The shape of the initial ellipsoid is specified by the three radii, $a_x$, $a_y$ and $a_z$ along its principal axes. Hence, the initial volume of the proto-halo is $V=(4\pi/3)a_x a_y a_z$.

The total simulation time $t_{\rm final}$ should be long enough for the halo to reach a virialised state. Normally, five free-fall times are sufficient for this equilibrium to occur (cf. later Figures 6~and~7). In the spherical collapse approximation, the free-fall time is given by
\begin{equation}
    t_{\rm ff}=\sqrt{\frac{3\pi}{32G\rho}},
\end{equation}
where $G$ is the gravitational constant and $\rho=M/V$ is the density of the sphere initially at rest. If we now choose the total simulation time $t_{\rm final}$ to be $5t_{\rm ff}$, the density becomes
\begin{equation}\label{eq:scalecondition}
    \frac{M}{V}=\frac{75\pi}{32G t_{\rm final}^2}.
\end{equation}
By virtue of the scale-freeness that gravity and ideal gas physics entail, the overall mass $M$ and simulation time $t_{\rm final}$ are irrelevant, in the sense that the solution is self-similar as long as the volume of the ellipsoid satisfies equation~\ref{eq:scalecondition}. For purely intuitive purposes, we choose $M=10^{12}\Msun$, roughly the mass of a typical Milky Way type halo, and $t_{\rm final}=14\cdot 10^9\rm\,yr$, roughly a Hubble time. Following these choices, the initial volume of the ellipsoid then implies a geometric mean radius of $\bar R=(a_x a_y a_z)^{1/3}\approx 306\rm\,kpc$. It then suffices to select the ratios of the radii $a_x$, $a_y$ and $a_z$ to solve for their absolute values. We chose a variety of axis ratios (see Table~\ref{tab:statisticalresults}), corresponding to both oblate and prolate spheroids, as well as a perfect sphere for verification purposes.

Having sampled the initial positions of the simulation particles, we are left to specify their velocities. We do so, by assigning to each particle (1) a random velocity component from an isotropic Maxwell distribution of 1D variance $\sigma^2$ and (2) a rotation component about the $z$-axis of fixed circular velocity $v_c$. In the spirit of the scale-invariance discussed before, it makes sense to express $\sigma$ and $v_c$ in dimensionless form. For the random dispersion, we do so through the parameter $\kappa$, defined by
\begin{equation}
    \sigma=\kappa \sqrt{\frac{GM}{\bar R}}.
    \label{equ:sigma}
\end{equation}
In turn, the rotation velocity $v_c$ is specified by the spin parameter $\lambda$.

The range of dispersion parameters $\kappa$ is limited by the fact that small dispersions ($\kappa\lesssim0.1$) lead to well-known radial collapse instabilities \citep{polyachenko_nonlinear_1981,huss_how_1999}, whereas for values $\kappa>1$ (corresponding to $E_{\rm kin}>|E_{\rm pot}|/2$ if $\lambda=0$) the halo would need to expand rather than collapse to reach virial equilibrium (if bound at all). For most of the control simulation, we use an intermediate value of $\kappa=0.2$, corresponding to $\sigma \approx23.7\rm\,km/s$ in the selected scales. To probe the dependency of the AM transfer on $\kappa$, we also ran a few simulations with $\kappa=0.15$ ($\sigma \approx17.8\rm\,km/s$) and $\kappa=0.25$ ($\sigma \approx29.6\rm\,km/s$).

The spin parameter is, by definition, positive and its maximum possible value in a uniform sphere is $(3/10)^{3/2}\approx0.1643$ (if $\kappa=0$). For most of our control runs we use $\lambda=0.03$ and $0.06$. The implied rotation velocities depend weakly on the choice of $\kappa$ (which alters the total energy of the system). It is about $v_c=5.9\rm\,km/s$ for $\lambda=0.03$ and $\kappa=0.2$. We also sample the spin parameter more finely ($\lambda=0.02,0.04,0.06,0.08,0.10$) for one specific ellipsoid.

Table~\ref{tab:statisticalresults} overviews the simulation parameters (shapes, spins, dispersions) used in the control simulations. To quantify the stochasticity coming from the random sampling of positions and dispersion velocities, each parameter configuration was repeated with five distinct seeds for the pseudo random number generator.

\subsection{Analysis of a single ellipsoidal collapse}
\label{ss:singlehalo}

\begin{figure*}
    \centering
    \includegraphics[width=\textwidth]{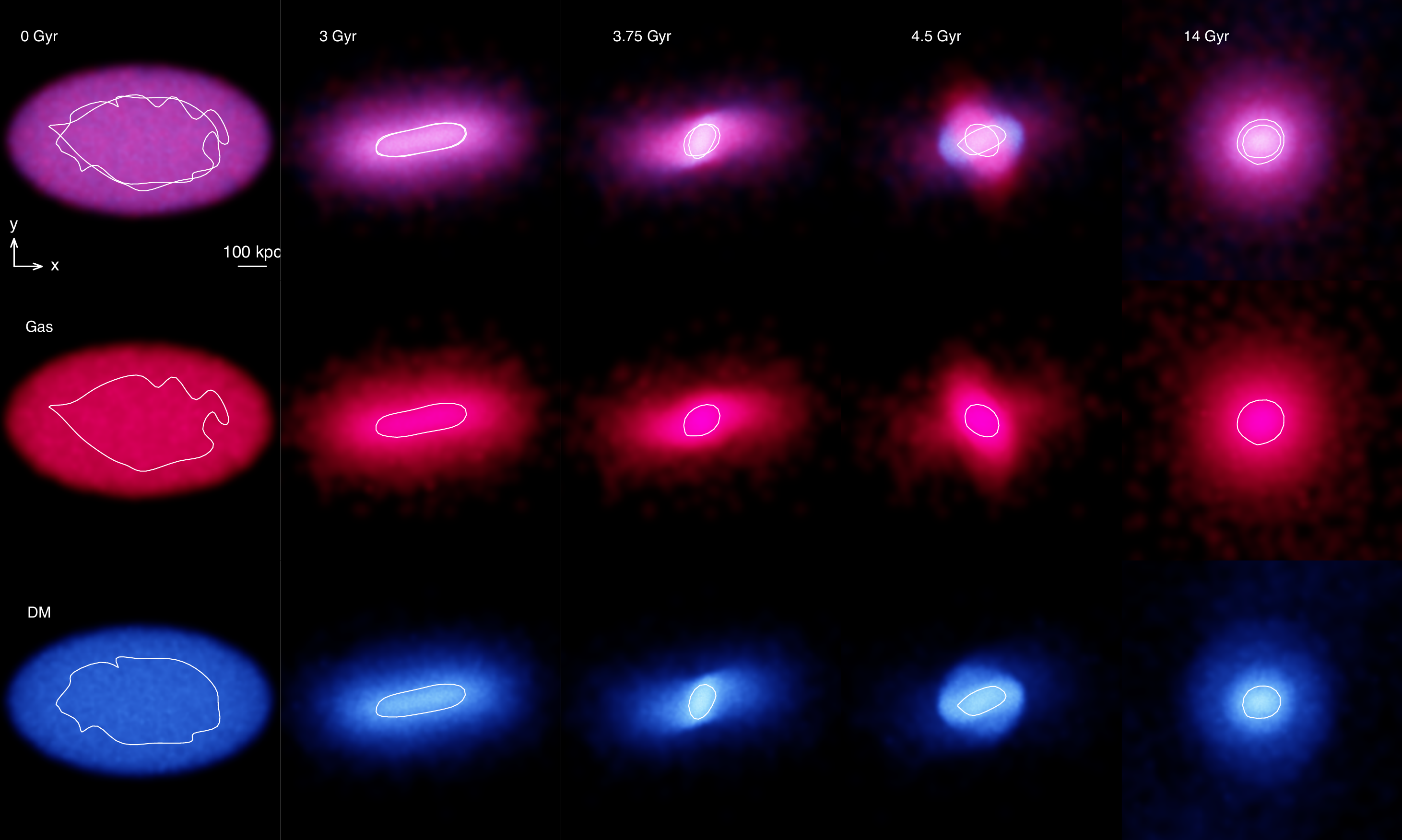}
    \caption{Five snapshots of a prolate halo rotating counterclockwise in the $xy$-plane. The set up of this halo is shown in Table~\ref{tab:statisticalresults} No.26. Each column from left to right represents the snapshot in 0, 3, 3.75, 4.5 and 14 Gyr, respectively. Each line from top to bottom represents total components, gas and DM respectively. The white contours include 50\% materials within it.}
    \label{fig:visulisation_halo}
\end{figure*}

Figure~\ref{fig:visulisation_halo} shows five time steps (snapshots) of one particular ellipsoidal top-hat with initial axes ratios $a_x:a_y:a_z=1.8:1.0:1.0$, dispersion $\kappa=0.2$ and spin $\lambda=0.06$. The system is shown in its projection on the $xy$-plane, which is orthogonal to the rotation axis. The rotation is anti-clockwise in this plane. As expected, the halo collapses more rapidly along its minor axes ($y$-axis and $z$-axis), such that the ellipticity increases rapidly for about one free-fall time ($\sim3\rm\,Gyr$). As the halo reaches its first collapse point (around $3.75\rm\,Gyr$), the inner half of the dark matter is slightly more compact than the corresponding inner gas mass due to the gas pressure acting against gravity. The contours in Figure~\ref{fig:visulisation_halo} show clearly that the DM is now spinning ahead of the gas, in qualitative agreement with Figure~\ref{fig:schematic}.

The DM keeps spinning ahead of the gas after the first shell-crossing (see $4\rm\,Gyr$). After a few more shell-crossings both components eventually settle into an almost spherically symmetric quasi-equilibrium state ($14\rm\,Gyr$).

\begin{figure}
    \centering
    \includegraphics[width=\columnwidth]{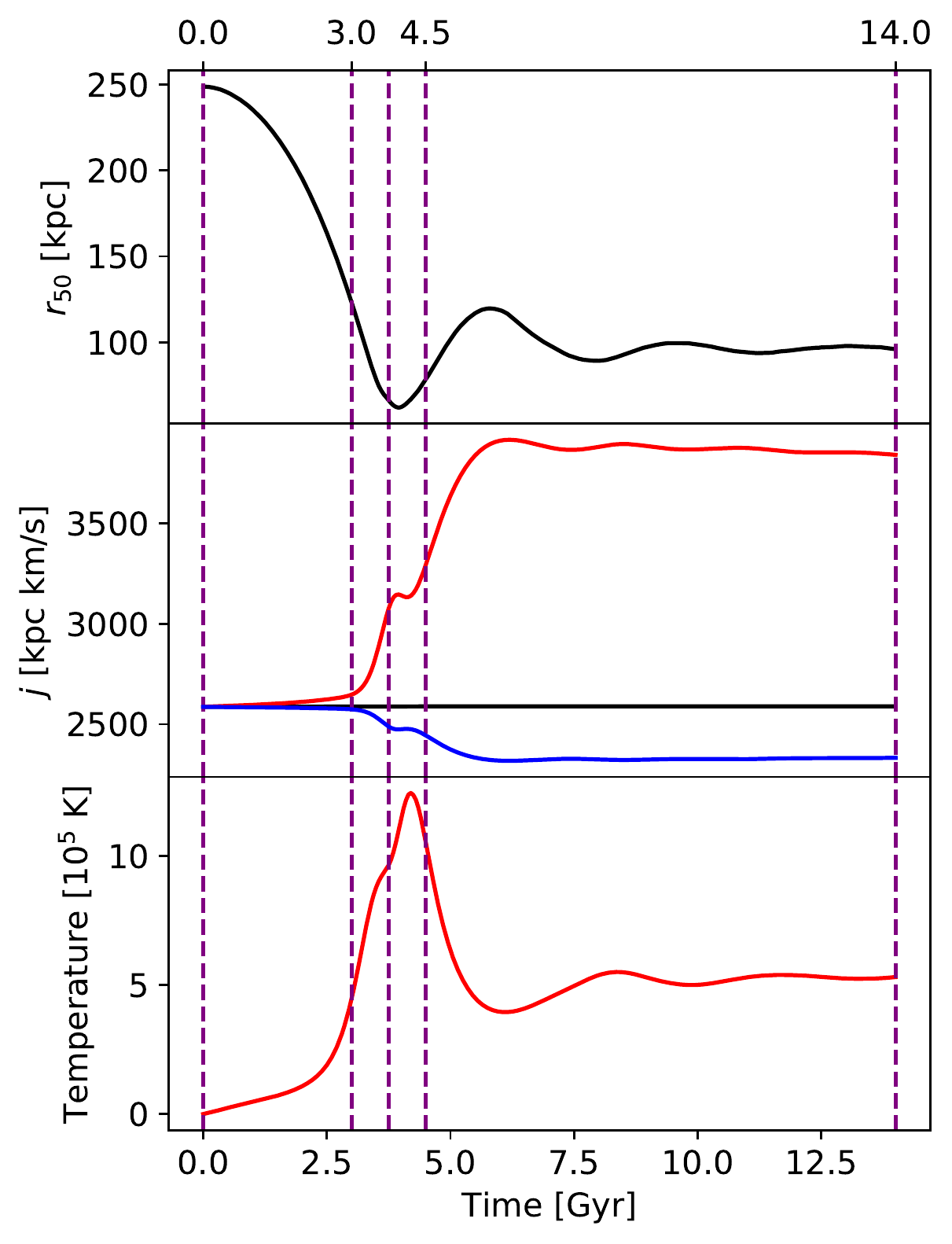}
    \caption{\textit{Top panel}: Half mass radius evolution of the halo. \textit{Middle panel}: sAM evolution of total components (black), gas (red) and DM (blue), respectively. \textit{Bottom panel}: gas temperature evolution of the halo gas. Five purple dashed lines show the positions of five snapshots in Figure~\ref{fig:visulisation_halo}.}
    \label{fig:CtrlSim_one_example}
\end{figure}

Figure~\ref{fig:CtrlSim_one_example} provides quantitative diagnostics for the evolution shown in Figure~\ref{fig:visulisation_halo}. The three panels in Figure~\ref{fig:CtrlSim_one_example} show the cosmic evolution of the half mass radius, the sAM and gas temperature, respectively (top to bottom). The vertical dashed lines indicate the times of the snapshots displayed in Figure~\ref{fig:visulisation_halo}. The evolution of the radius clearly reveals the initial collapse, paralleled by a sharp rise in gas temperature. This collapse is followed by a virialisation phase, characterized by a series of damped oscillations in size and temperature. Towards the end of the simulation, all diagnostic curves become constant as the halo reaches a quasi-equilibrium state.

Since the halo is simulated as an isolated system, its total AM (about the centre of mass) remains constant, as indicated by the black line in the middle panel of Figure~\ref{fig:CtrlSim_one_example}. However, AM is transferred from the DM to the gas at the final collapse stage (3--3.7~Gyr), as expected from the torque considerations schematised in Figure~\ref{fig:schematic}. Perhaps less expected is the second, similarly important transfer of AM during the post-collapse expansion phase (4-6~Gyr). The physical reason for this second transfer is the same: the more concentrated DM ellipsoid rotates ahead of the gas ellipsoid, thus causing a net torque that spins up the gas and spins down the DM, proportionally to their masses. This non-trivial time-lime of AM transfer makes attempts to model this full process analytically very challenging. In the following we will instead follow a more heuristic way to quantitatively model the AM transferred in the control simulations.

\subsection{Statistical analysis of all control runs}

Having illustrated how AM is transferred from DM to gas during an ellipsoidal top-hat collapse, we shall now quantify this process as a function of the initial conditions. Explicitly, we will quantify the AM acquired by the gas as a function of the overall spin parameter, the initial geometry, and the collapse factor. The latter is controlled via the dispersion parameter $\kappa$ (see Section~\ref{ss:ctl_setup}). Table~\ref{tab:statisticalresults} lists the initial conditions considered in this analysis.

Let us first elaborate on how the AM transfer between DM and gas is best quantified. So far, we have followed the literature and relied on \jgas/\jDM ($\equiv\lambda_{\rm gas}/\lambda{\rm DM}$) as a way of measuring this transfer. However, this parameter has the disadvantage that its denominator may get close to zero, making it ill-defined in some cases. Moreover, the higher the initial value of \jgas and \jDM, the smaller the variation in \jgas/\jDM for a fixed amount of AM transfer per unit mass. For our analysis, it is more appropriate to consider directly the AM, $\Delta J$, acquired by the gas. This quantity can be computed via
\begin{equation}
    \Delta J = J_{\rm gas}(t_f)-J_{\rm gas}(t_i)=-[J_{\rm DM}(t_f)-J_{\rm DM}(t_i)].
    \label{equ:ctl_totalAMtransfer}
\end{equation}
where $t_i$ and $t_f$ are the initial and final time, respectively. Of course, absolute AM values scale with the overall mass of the system. To remove this trivial scale dependence, we need to form a dimensionless {\it spin transfer parameter} $\Delta \lambda$. In analogy to the Peebles parameter (equation~\ref{equ:lambda_P}), we set
\begin{equation}
    \Delta \lambda=\frac{\Delta J |E|^{1/2}}{GM^{5/2}}.
    \label{equ:deltalambda_ctl}
\end{equation}
Note that $E$ and $M$ here are the energy and mass of the whole system respectively, thus they are constant during the simulation time. We stress that $\Delta \lambda$ is {\it not} measuring a change in the spin parameter $\lambda$ from time $t_i$ to $t_f$, but it is a normalised measure of how much AM gets transferred internally from DM to gas. In fact, the (DM+gas) spin parameter $\lambda$ remains constant in the isolated control runs.

\subsubsection{Spin transfer as a function of overall spin}

\begin{figure}
    \centering
    \includegraphics[width=\columnwidth]{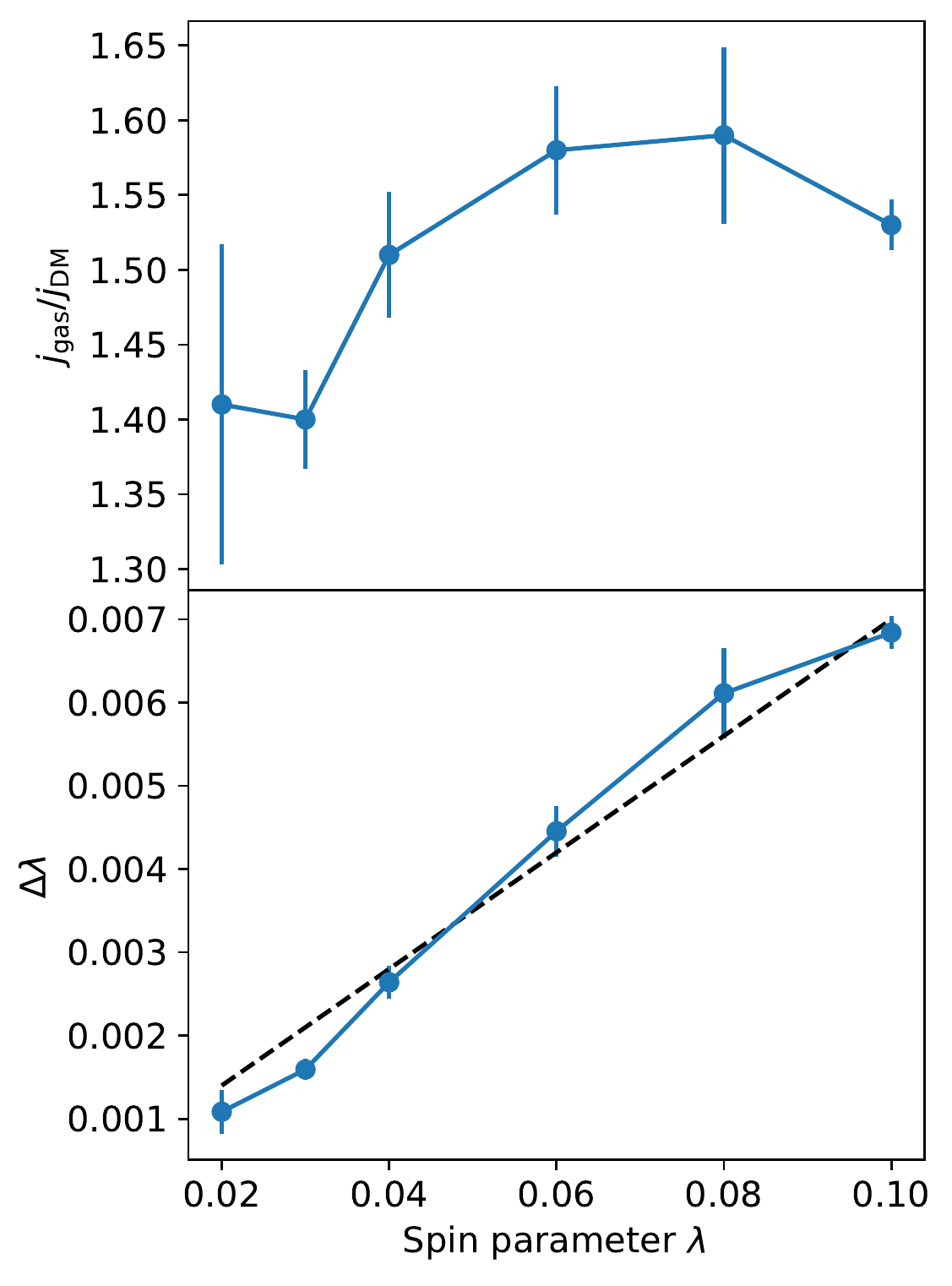}
    \caption{{\it Top panel:} $\lambda$-\jgas/\jDM relation from the control simulations, which does not increase monotonically. {\it Bottom panel:} $\lambda$-$\Delta \lambda$ relation from the control simulations. $\Delta \lambda $ is a normalised measure of how much AM was transferred internally from DM to gas, defined by equation~\ref{equ:deltalambda_ctl}. The $\lambda-\Delta \lambda$ relation appears to be approximately linear, as shown by the black dashed line for illustrative purposes only. The error bars show the standard deviation of five random realisations of each set of initial simulation parameters.}
    \label{fig:spin_comp}
\end{figure}

Figure~\ref{fig:spin_comp} shows both \jgas/\jDM and $\Delta\lambda$ as a function of $\lambda$, at fixed initial axis ratio (1.8:1:1) and dispersion ($\kappa=0.2$). While \jgas/\jDM varies non-linearly (even non-monotonically) with $\lambda$, the spin transfer parameter $\Delta\lambda$ is nearly proportional to $\lambda$. This proportionality is expected from the schematic picture of Figure~\ref{fig:schematic}, whereby the torque is predicted to be proportional to the angle between the major axes of the DM and gas, which itself is expected to be proportional to $\lambda$. This explanation naturally fails for large angles ($\gtrsim45^\circ$), which may explain the slight bend in the $\lambda$--$\Delta\lambda$ relation at the highest spins (right-most point).

The comparison in Figure~\ref{fig:spin_comp} underscores the advantage of $\Delta\lambda$ over \jgas/\jDM. We will stick with $\Delta\lambda$ for most of the following analysis, but will return to \jgas/\jDM in Section~\ref{s:discussion}, when attempting to explain the mean AM excess in gas in non-radiative cosmological simulations.

\subsubsection{Spin transfer as a function of halo geometry}

A look at Table~\ref{tab:statisticalresults} reveals that the initial axis ratios, including the type of ellipsoid (prolate versus oblate), strongly affect the AM transfer from DM to gas. Numerically, we find that, at fixed $\lambda$ and $\kappa$, the spin transfer parameter $\Delta\lambda$ increases monotonically with the dimensionless {\it geometry parameter}
\begin{equation}
    g=\frac{a_x^2-a_y^2}{a_z^2},
    \label{equ:geoparameter}
\end{equation}
where $a_x$ and $a_y$ are the major and minor axes, respectively, in the plane orthogonal to the total AM vector. The relation between $g$ and $\Delta\lambda$, at fixed $\lambda=0.03$ and $\kappa=0.2$, is shown in Figure~\ref{fig:delta_twopara} (left). Clearly, $\Delta\lambda=0$ if $g=0$, no matter if the halo is oblate or prolate, highlighting that no AM can be transferred if the halo is rotationally symmetric about the rotation axis. The relation is not quite proportional, but a linear approximation (with zero offset) will suffice for our purposes. The normalisation by $a_z^2$ in the definition of $g$ makes oblate and prolate haloes fall roughly on the same relation.

\subsubsection{Spin transfer as a function of collapse factor}

Finally, we consider the variation of $\Delta\lambda$ as a function of the collapse factor
\begin{equation}
    c=\frac{r_{50}(t_i)}{r_{50}(t_f)},
    \label{equ:collapsefac_ctl}
\end{equation}
where $r_{50}$ is the half-mass radius. We can approximately control the value of $c$ via the dispersion parameter $\kappa$, which controls the dynamic support against gravitational collapse.

\begin{figure*}
    \centering
    \includegraphics[width=\textwidth]{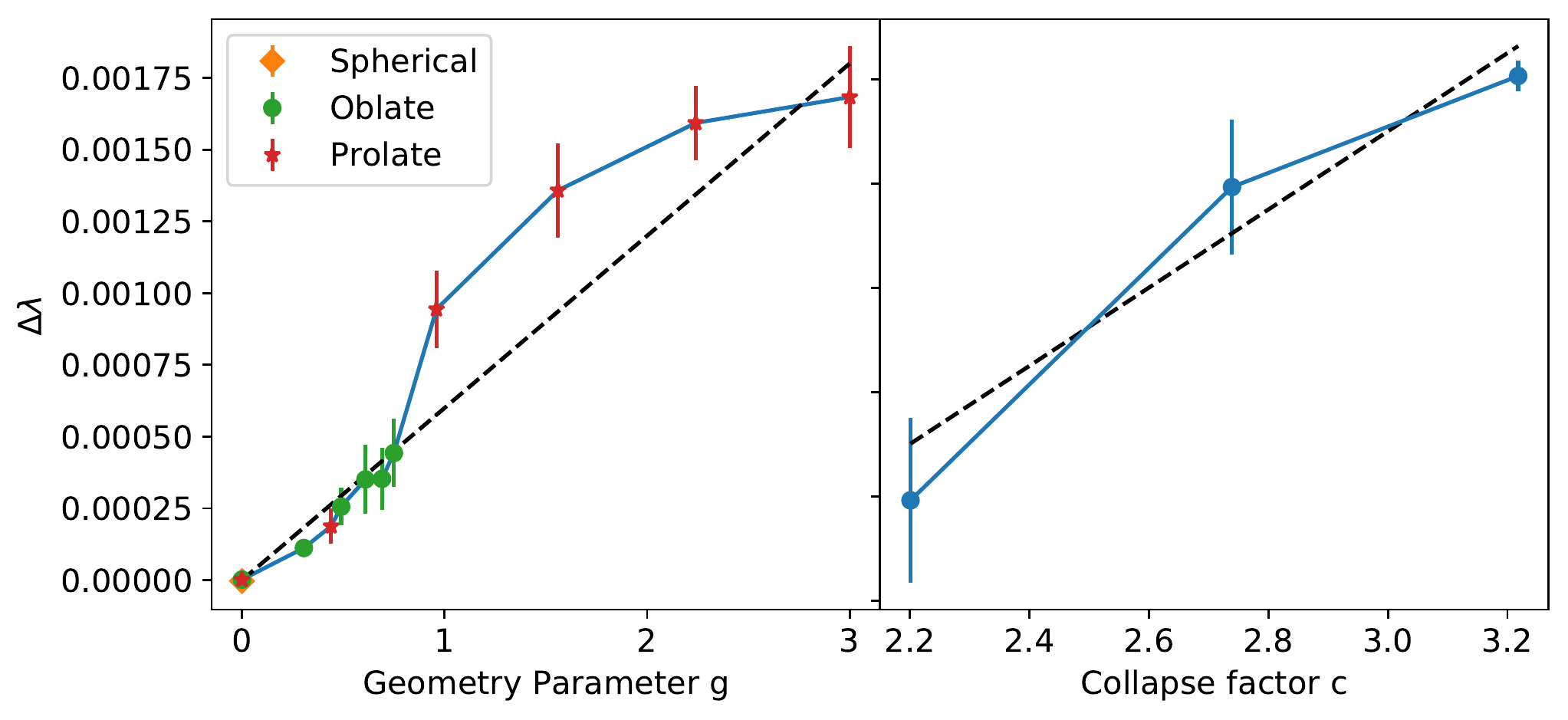}
    \caption{{\it Left panel:} Geometry parameter-$\Delta \lambda$ relation. The geometry parameter is defined as equation~\ref{equ:geoparameter}. Different symbols represent spherical, oblate and prolate shapes. More AM transfers from DM to gas in haloes with a larger asymmetry in the plane of rotation. {\it Right panel:} Collapse factor-$\Delta \lambda$ relation. The collapse factor is defined by equation~\ref{equ:collapsefac_ctl}. More AM transfers from DM to gas in haloes with a larger collapse factor. In both panels, the error bars show the standard deviations of five random realisations of the simulations; and the black dashed lines show linear models for purely illustrative purposes.}
    \label{fig:delta_twopara}
\end{figure*}

Since the ellipticity of a halo increases as the halo contracts, we would expect that higher collapse factors lead to higher asymmetries near the collapse point and thus more AM transfer. This is shown in Figure~\ref{fig:delta_twopara} (right), at fixed geometry (1.8:1:1) and $\lambda=0.03$. The relationship is again roughly proportional, although the considered range of $c$ is quite small, due to the physical limits of $\kappa$ discussed in Section~\ref{ss:ctl_setup}.

 \begin{table*}
    \centering
    \begin{tabular}{ccccccccccccc}
        \hline
        No. & $a_x:a_y:a_z$ & $\lambda$ & $\kappa$ & $c$ & $\Delta c$ & $j_{\rm gas,ini}$ & $j_{\rm gas,fin}$ & $j_{\rm DM,ini}$ & $j_{\rm DM,fin}$ & \jgas/\jDM & $\Delta$(\jgas/\jDM) & $\Delta \lambda$\\
        \hline
        1 & 1.0:1.0:1.0 & 0.03 & 0.20 & 3.108 & 0.015 & 1288 & 1287 & 1288 & 1288 & 1.00 & 0.002 & 0.00000\\
2 & 1.0:1.0:1.0 & 0.06 & 0.20 & 2.937 & 0.010 & 2586 & 2587 & 2586 & 2587 & 1.00 & 0.002 & 0.00000\\
3 & 1.0:1.0:1.8 & 0.03 & 0.20 & 2.774 & 0.024 & 1289 & 1290 & 1289 & 1289 & 1.00 & 0.001 & 0.00000\\
4 & 1.0:1.0:1.8 & 0.06 & 0.20 & 2.673 & 0.012 & 2594 & 2594 & 2594 & 2594 & 1.00 & 0.000 & 0.00000\\
5 & 1.2:1.0:1.2 & 0.03 & 0.20 & 3.031 & 0.015 & 1288 & 1317 & 1288 & 1283 & 1.03 & 0.003 & 0.00011\\
6 & 1.2:1.0:1.2 & 0.06 & 0.20 & 2.854 & 0.007 & 2587 & 2656 & 2587 & 2575 & 1.03 & 0.006 & 0.00027\\
7 & 1.4:1.0:1.4 & 0.03 & 0.20 & 2.903 & 0.020 & 1288 & 1353 & 1288 & 1275 & 1.06 & 0.016 & 0.00026\\
8 & 1.4:1.0:1.4 & 0.06 & 0.20 & 2.805 & 0.007 & 2587 & 2737 & 2587 & 2557 & 1.07 & 0.017 & 0.00059\\
9 & 1.6:1.0:1.6 & 0.03 & 0.20 & 2.876 & 0.015 & 1292 & 1381 & 1292 & 1274 & 1.08 & 0.029 & 0.00035\\
10 & 1.6:1.0:1.6 & 0.06 & 0.20 & 2.834 & 0.012 & 2594 & 2817 & 2594 & 2549 & 1.11 & 0.012 & 0.00087\\
11 & 1.8:1.0:1.8 & 0.03 & 0.20 & 2.912 & 0.015 & 1301 & 1391 & 1301 & 1283 & 1.08 & 0.027 & 0.00035\\
12 & 1.8:1.0:1.8 & 0.06 & 0.20 & 2.913 & 0.018 & 2614 & 2901 & 2614 & 2553 & 1.14 & 0.014 & 0.00113\\
13 & 2.0:1.0:2.0 & 0.03 & 0.20 & 2.962 & 0.011 & 1313 & 1427 & 1313 & 1289 & 1.11 & 0.031 & 0.00044\\
14 & 2.0:1.0:2.0 & 0.06 & 0.20 & 2.975 & 0.004 & 2637 & 2954 & 2638 & 2570 & 1.15 & 0.016 & 0.00123\\
15 & 1.2:1.0:1.0 & 0.03 & 0.20 & 2.979 & 0.016 & 1288 & 1336 & 1288 & 1278 & 1.05 & 0.014 & 0.00019\\
16 & 1.2:1.0:1.0 & 0.06 & 0.20 & 2.827 & 0.015 & 2586 & 2692 & 2586 & 2566 & 1.05 & 0.009 & 0.00041\\
17 & 1.4:1.0:1.0 & 0.03 & 0.20 & 2.794 & 0.005 & 1288 & 1527 & 1289 & 1240 & 1.23 & 0.034 & 0.00094\\
18 & 1.4:1.0:1.0 & 0.06 & 0.20 & 2.797 & 0.009 & 2586 & 3148 & 2586 & 2473 & 1.27 & 0.025 & 0.00221\\
19 & 1.6:1.0:1.0 & 0.03 & 0.20 & 2.819 & 0.015 & 1287 & 1630 & 1288 & 1219 & 1.34 & 0.043 & 0.00135\\
20 & 1.6:1.0:1.0 & 0.06 & 0.20 & 2.766 & 0.010 & 2583 & 3480 & 2584 & 2403 & 1.45 & 0.010 & 0.00353\\
21 & 1.8:1.0:1.0 & 0.02 & 0.20 & 2.748 & 0.014 & 858 & 1132 & 858 & 803 & 1.41 & 0.107 & 0.00108\\
22 & 1.8:1.0:1.0 & 0.03 & 0.15 & 3.218 & 0.041 & 1278 & 1731 & 1278 & 1186 & 1.46 & 0.010 & 0.00181\\
23 & 1.8:1.0:1.0 & 0.03 & 0.20 & 2.739 & 0.020 & 1288 & 1690 & 1288 & 1207 & 1.40 & 0.033 & 0.00159\\
24 & 1.8:1.0:1.0 & 0.03 & 0.25 & 2.201 & 0.007 & 1300 & 1553 & 1300 & 1249 & 1.24 & 0.041 & 0.00099\\
25 & 1.8:1.0:1.0 & 0.04 & 0.20 & 2.706 & 0.003 & 1718 & 2385 & 1718 & 1585 & 1.51 & 0.042 & 0.00263\\
26 & 1.8:1.0:1.0 & 0.06 & 0.20 & 2.591 & 0.003 & 2583 & 3711 & 2583 & 2355 & 1.58 & 0.043 & 0.00444\\
27 & 1.8:1.0:1.0 & 0.08 & 0.20 & 2.471 & 0.003 & 3453 & 5006 & 3454 & 3140 & 1.59 & 0.059 & 0.00610\\
28 & 1.8:1.0:1.0 & 0.10 & 0.20 & 2.394 & 0.006 & 4333 & 6077 & 4333 & 3980 & 1.53 & 0.017 & 0.00683\\
29 & 1.8:1.8:1.0 & 0.03 & 0.20 & 2.923 & 0.008 & 1301 & 1301 & 1301 & 1300 & 1.00 & 0.001 & 0.00000\\
30 & 1.8:1.8:1.0 & 0.06 & 0.20 & 2.926 & 0.008 & 2608 & 2612 & 2608 & 2609 & 1.00 & 0.001 & 0.00001\\
31 & 2.0:1.0:1.0 & 0.03 & 0.20 & 2.588 & 0.014 & 1288 & 1713 & 1288 & 1203 & 1.42 & 0.048 & 0.00168\\
32 & 2.0:1.0:1.0 & 0.06 & 0.20 & 2.495 & 0.007 & 2583 & 3661 & 2583 & 2365 & 1.55 & 0.031 & 0.00425\\
\\[-2ex]
        \hline
    \end{tabular}
    \caption{Initial conditions and results of sAM of all the control simulations. All the haloes rotate along $z$ axis. Column (1): Simulation number; (2) the $a_x:a_y:a_z$ ratio of the spherical/elliptical (proto-)haloes (3) Initial Peebles spin parameter; (4) velocity dispersion factor; (5) collapse factor defined as equation~\ref{equ:collapsefac_ctl}; (6) standard deviation of collapse factor; (7) sAM of gas in the initial condition; (8) sAM of gas in the final snapshot; (9) sAM of DM in the initial condition; (10) sAM of DM in the final snapshot; (11) ratio of sAM of gas to DM; (12) standard deviation of sAM ratio. (13) spin transfer parameter $\Delta \lambda$, defined as equation~\ref{equ:deltalambda_ctl}.}
    \label{tab:statisticalresults}
\end{table*}

\begin{figure}
    \centering
    \includegraphics[width=\columnwidth]{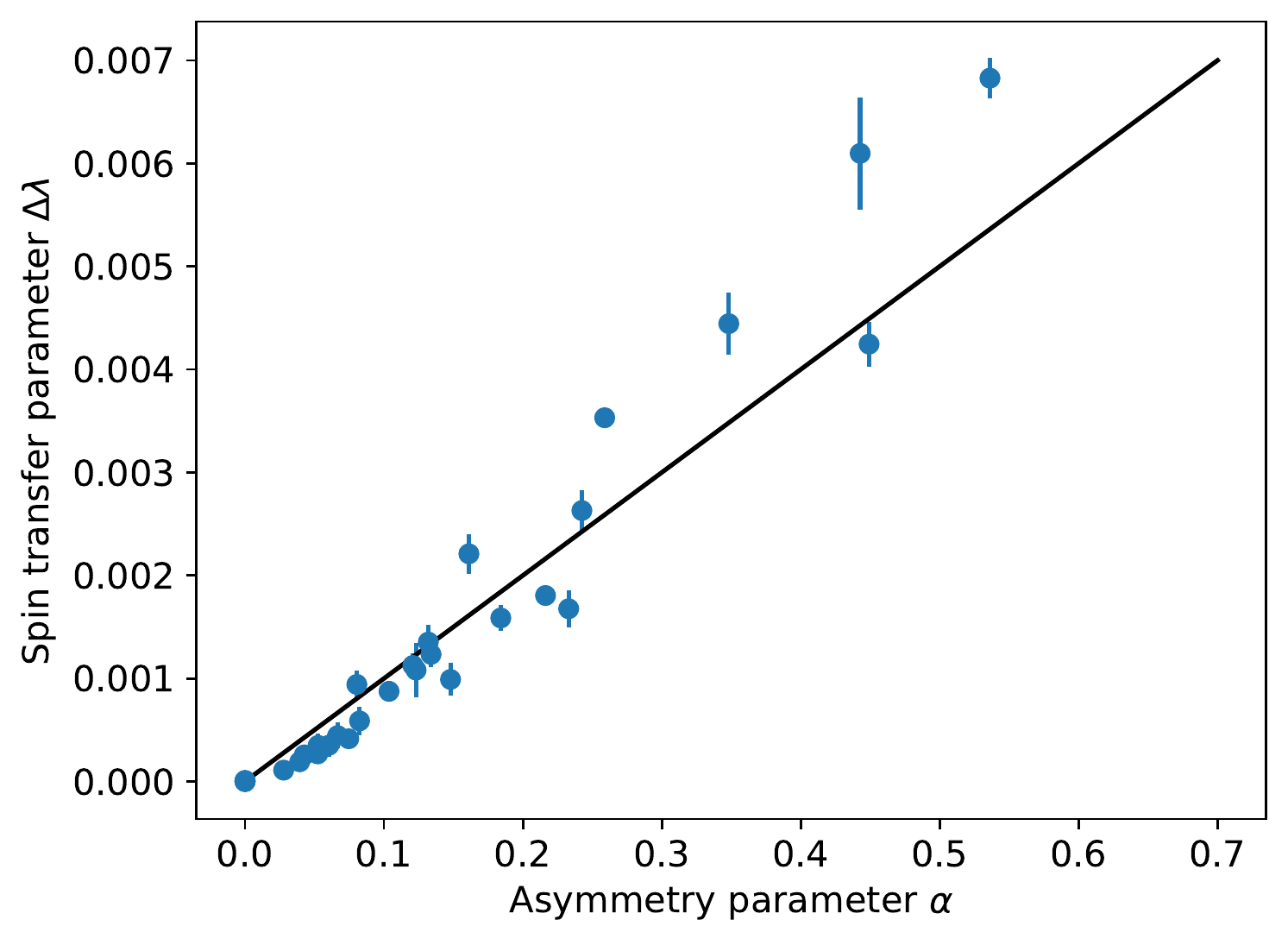}
    \caption{$\alpha-\Delta \lambda$ relation from the control simulations. The black line shows the fit of equation~\ref{equ:deltalambdaalpha}, with $k=0.01$. The errorbars are the standard deviations of 5 random seeds.}
    \label{fig:alpha_delta_ctl}
\end{figure}

\subsubsection{Full model of the spin transfer}

We have shown that in simulations of an ellipsoidal top-hat collapse, the AM transferred from DM to gas, as measured by $\Delta\lambda$, varies approximately proportionally with the spin parameter $\lambda$, the initial geometry factor $g$ and the collapse factor $c$. To the extent that these three variations are separable, we then expect that $\Delta\lambda$ is also proportional to the product of these three parameters. This leads us to define an {\it asymmetry parameter} (for lack of a better name),
\begin{equation}
    \alpha = g\,\lambda\,c = \frac{a_x^2-a_y^2}{a_z^2}\times\lambda\times\frac{r_{50}(t_i)}{r_{50}(t_f)}.
    \label{equ:asym_para}
\end{equation}
We then fit the linear model
\begin{equation}
    \Delta \lambda = k\alpha
    \label{equ:deltalambdaalpha}
\end{equation}
using a $\chi^2$-minimisation, resulting in $k=0.01\pm 0.002$. The uncertainty range represents the standard deviation inferred from $10^4$ bootstrapping iterations, where each iteration used a random subsample of 80 simulations from our 160 runs (32 different initial conditions with 5 random seeds each).

Figure~\ref{fig:alpha_delta_ctl} shows this linear fit alongside the simulation data. The comparison demonstrates that the linear model approximately captures the systematic dependence of $\Delta \lambda$ on the initial parameters of the ellipsoidal top-hat collapse. Unlike in Figures \ref{fig:spin_comp} and \ref{fig:delta_twopara}, the simulation data shown in Figure~\ref{fig:alpha_delta_ctl} also includes cases, where multiple initial parameters were varied simultaneously (see Table \ref{tab:statisticalresults}).

In a strict statistical sense, the linear model does not fully describe the simulation data (as suggested by a reduced $\chi^2$-value of about 4 for the binned data). However,  compared to the large halo-to-halo scatter of $\Delta \lambda$ in each $\alpha$ bin ($\sim 0.7$ dex, see Section~\ref{s:discussion}), a linear model largely suffices to describe the systematics.

\subsection{Limitation of control simulations}
Top-hat models provide some insight into AM transfer during halo collapse. However, these toy models have some limitations: (1) The top-hat (proto-)haloes assume that the density is uniform, and all the particles collapse at the same time. But real haloes have density fluctuations, where the denser regions collapse in advance and form substructures. This asynchronous collapse alters the local torques. (2) We set up the spin vectors of gas and DM to be perfectly aligned, but cosmological simulations have found misalignment between gas and DM, which are $\sim 20^{\circ}-30^{\circ}$ \citep[e.g.][]{van_den_bosch_angular_2002,sharma_origin_2012,bett_angular_2010} on average. This misalignment may induce torques inside the haloes, changing the amount of AM transfer. (3) We have only considered spheroidal (i.e. spherical, oblate and prolate) shapes. However, in cosmological simulations, haloes are often triaxial or irregular. The complex of shape may affect the AM transfer. (4) We do not consider environmental effects and external torques outside the haloes (post turn-around). External torques may affect gas and DM differently and change their final spin ratio. We will discuss such effects in the cosmological simulations in Section~\ref{s:discussion}.

\section{Discussion}
\label{s:discussion}

Our idealised simulations of the isolated ellipsoidal top-hat collapse with DM and gas have led to a heuristic model (equations \ref{equ:asym_para} and \ref{equ:deltalambdaalpha}) that accurately describes the AM transfer between the two substances. This model quantifies the intuitive physical process schematised in Figure~\ref{fig:schematic}. The purpose of this section is to determine whether and to what extent this simple model can also explain the AM differences in gas and DM seen in the adiabatic cosmological simulation from SURFS (Section \ref{s:cosmosims}).

To analyse the SURFS haloes consistently with the ellipsoidal top-hats, we must trace their particles back in time from $z=0$ to the beginning of the collapse. In an expanding universe, this corresponds to the so-called turn-around point, where the self-gravity of a density peak starts to dominate over the dark energy pressure. We determine this turn-around point individually for each halo by finding the snapshot, where the {\it physical} mass-weighted rms-radius of the traced-back particles reaches its maximum. By definition, the velocity of the rms-radius vanishes at this point, consistent with the initial conditions of our ellipsoidal top-hat simulations. The turn-around redshift $z_{\rm turn}$ depends on the halo mass and concentration \citep[e.g.][]{ludlow_mass-concentration-redshift_2016}. Its average value for our selection ($M\geqslant10^{12}h^{-1}\rm{M_{\odot}}$) is about $z_{\rm turn}\approx2$. 

As emphasised in Section \ref{s:cosmosims}, the AM ratio between gas and DM is close to unity at $z_{\rm turn}$, but increases as the halo contracts and virialises. This was indeed the motivation to consider specifically the collapse phase in Section~\ref{s:controlsims}.

To apply the model of equations \ref{equ:asym_para} and \ref{equ:deltalambdaalpha} to the SURFS haloes, we must first extend the definitions of the asymmetry parameter $\alpha$ and the spin transfer parameter $\Delta\lambda$ to the case of cosmological simulations. We will do so, while making sure that the extended definitions reduce to the previous ones, if applied to an isolated ellipsoid.

For most of this section, the initial time $t_i$ corresponds to the cosmic time at $z_{\rm turn}$ and $t_f$ corresponds to the current age of the universe at $z=0$.

\subsection{Model extension to cosmological simulations}

\subsubsection{Asymmetry parameter in cosmological simulations}

Computing the asymmetry parameter $\alpha$, defined in equation~\ref{equ:asym_para}, requires extended definitions of its three factors: the initial geometry factor $g$, the collapse factor $c$ and the spin parameter $\lambda$.

To extend the geometry factor to the case of a general particle distribution without a uniquely defined boundary, we substitute the three axes $a_x$, $a_y$ and $a_z$ of the initial ellipsoid for second moments,
\begin{equation}\label{equ:geoparameter2}
	g = \frac{\avg{\tilde{x}^2}-\avg{\tilde{y}^2}}{\avg{\tilde{z}^2}},
\end{equation}
evaluated at the initial time $(t_i)$, using {\it all} the particles that constitute a halo at $t_f$ ($z=0$). Brackets $\avg{...}$ represent mass-weighted averages over all these particles. The tilde symbols refer to the proper coordinates, defined separately for each halo. These coordinates are such that the centre of mass lies at the origin, i.e.~$\avg{\tilde{x}}=\avg{\tilde{y}}=\avg{\tilde{z}}=0$. The $\tilde{z}$-axis points along with the total angular momentum $\mathbf{J}(t_i)$ of the system. In the orthogonal plane, the $\tilde{x}$-axis is defined as the direction that maximises $\avg{\tilde{x}^2}$, and thus the perpendicular $\tilde{y}$-axis minimises $\avg{\tilde{y}^2}$. In practice these two directions are found by diagonalising the moment matrix
\begin{equation}
\mathbf{M}=\begin{pmatrix}
    \avg{\tilde{x}^2} & \avg{\tilde{x}\tilde{y}} \\
    \avg{\tilde{x}\tilde{y}} & \avg{\tilde{y}^2}
\end{pmatrix}.
\end{equation}
Note that this extended definition of $g$ (equation~\ref{equ:geoparameter2}) is identical to the earlier definition (equation~\ref{equ:geoparameter}) in the special case of a uniform ellipsoid.

Similarly, we can extend the definition of the collapse factor (equation~\ref{equ:collapsefac_ctl}) in terms of second moments:
\begin{equation}
	\label{eq:ccosmo}
    c = \sqrt{\frac{\avg{\tilde r^2(t_i)}}{\avg{\tilde r^2(t_f)}}},
\end{equation}
where $\tilde r^2=\tilde x^2+\tilde y^2+\tilde z^2$. This ratio of rms radii is not strictly equal to the ratio of half-mass radii, but the difference is insignificant in the context of this analysis of traced-back halo particles. We here prefer the use of rms radii for it allows us to express both $g$ and $c$ in terms of second moments.

As for the spin parameter $\lambda$, we move from Peeble's convention (equation~\ref{equ:lambda_P}) to Bullock's convention (equation~\ref{equ:lambda_B}), noting that their numerical values are similar (within 20 percent) for virialised haloes. Since for the traced-back particles, we do not have direct access to the virial radius and velocity used in Bullock's formulation, we express $\lambda'$ in terms of mass using the standard virial scaling relations, such that \citep[see][equation 4]{obreschkow_low_2015}.
\begin{equation}
    \lambda' = \frac{j H^{1/3}(z)\Delta_c^{1/6}(z)}{(2GM_{\rm halo})^{2/3}},
    \label{equ:spincosmo}
\end{equation}
where $H(z)$ is the Hubble `constant' that can be expanded as $H(z)=H_0E(z)$, where $H_0$ is the local Hubble constant and $E(z)=(\Omega_m(1+z)^3+\Omega_\Lambda)^{1/2}$ in a flat universe with a cosmological constant $\Lambda$. The scalar function $\Delta_c(z)$ is the over-density factor of virialized (spherical) density peaks relative to the mean density of the universe. Following \cite{bryan_statistical_1998}, it can be approximated as $\Delta_c(z)=18\pi^2+82\Omega_{\Lambda}E^{-2}(z)-39\Omega_{\Lambda}^2E^{-4}(z)$.

For equation~\ref{equ:spincosmo} to apply, the redshift $z$ must correspond to the instant at which $M$ is the virial mass. Since we are tracing back the particles of (approximately) virialised haloes at $z=0$, the spin parameter of these systems at {\it any} earlier time, should therefore still be computed while setting $z=0$ in equation \ref{equ:spincosmo}. The spin most relevant to the internal AM transfer from DM to gas during the collapse is arguably the spin at the onset of the collapse. We therefore evaluate $\lambda'$ using $j(t_i)$, but note that $\lambda'$ remains roughly preserved during the collapse, such that the following results would not significantly change if we used $j(t_f)$ instead.

In summary, we compute the asymmetry parameter $\alpha$ of the haloes in the cosmological simulation using equation~\ref{equ:asym_para}, but with the extended definitions of $g$, $c$ and $\lambda\rightarrow\lambda'$ in equations~\ref{equ:geoparameter2}, \ref{eq:ccosmo} and \ref{equ:spincosmo}.

\subsubsection{Spin transfer parameter in cosmological simulations}

In isolated haloes, such in the control simulations, the AM $\Delta J$ transferred {\it internally} from the DM to the gas over the time interval $[t_i,t_f]$, is equal to the change in gas AM during that interval (see equation~\ref{equ:ctl_totalAMtransfer}). However, if interacting with an environment, the gas can acquire (or lose) additional AM $\Delta J_{\rm gas,ext}$ from external torques, such that
\begin{equation}
    \Delta J_{\rm gas,ext}+\Delta J = J_{\rm gas}(t_f)-J_{\rm gas}(t_i).
    \label{equ:ctl_totalAMtransfer2}
\end{equation}
In principle, this is a vector equation, but we make the approximation that the AM axis remains constant in order to arrive at a simplified analytical model of practical usability. To gauge the errors induced by this approximation, we have explicitly computed the change in the gas spin direction between the SURFS haloes at $z=0$ and their corresponding Lagrangian regions at turn-around (see also \citealp{power_seeking_2013}; not to be confused with the main progenitor haloes; see \citealp{contreras_angular_2017} for this case). The median change in the gas spin direction is $31^\circ$, corresponding to a $14\%$ error, if the initial spin is projected on the final spin axis and vice versa. This is an acceptable systematic error in light of the statistical scatter of about 0.7 dex in $\xspingas/\xspinDM$ (see Figure \ref{fig:predictlambda}). Upon further assuming that external gravitational torque fields are identical for the gas and DM, their externally acquired AM is proportional to their mass. Hence, $\Delta J_{\rm gas,ext}=\f\Delta J_{\rm tot,ext}$, where $\f$ is the baryon fraction of the halo and $\Delta J_{\rm tot,ext}$ is the total externally acquired AM in the time interval of interest, i.e. $\Delta J_{\rm tot,ext}=J_{\rm tot}(t_f)-J_{\rm tot}(t_i)=J_{\rm DM}(t_f)+J_{\rm gas}(t_f)-[J_{\rm DM}(t_i)+J_{\rm gas}(t_i)]$. With these identities we can recast equation~\ref{equ:ctl_totalAMtransfer2} to
\begin{equation}
    \Delta J = (1\!-\!\f)[J_{\rm gas}(t_f)\!-\!J_{\rm gas}(t_i)]-\f[J_{\rm DM}(t_f)\!-\!J_{\rm DM}(t_i)].
    \label{equ:ctl_totalAMtransfer3}
\end{equation}
It is sometimes convenient to express this in the specific form $\Delta j\equiv\Delta J/M$, where $M=M_{\rm gas}+M_{\rm DM}$ is the total mass of the halo. In this form, equation \ref{equ:ctl_totalAMtransfer3} further simplifies to
\begin{equation}
	\Delta j = (\f-\f^2)(\Delta\jg-\Delta\jd),
    \label{equ:ctl_totalAMtransfer4}
\end{equation}
where $\Delta\jg=\jg(t_f)-\jg(t_i)$ and $\Delta\jd=\jd(t_f)-\jd(t_i)$.

In analogy to the spin parameter definition in equation~\ref{equ:spincosmo}, we can now express $\Delta j$ in dimensionless form via
\begin{equation}
    \Delta\lambda = \frac{\Delta j\,H^{1/3}(z)\Delta_c^{1/6}(z)}{(2GM_{\rm halo})^{2/3}}.
    \label{equ:spintrans_cosmo}
\end{equation}
Again, this equation needs to be evaluated using $z=0$, since this is the moment at which the halo mass $M_{\rm halo}$ (i.e. the sum of all particle masses in the FOF-group) is expected to be approximately equal to the virial mass.

Note that the numerical values of equations~\ref{equ:deltalambda_ctl} and \ref{equ:spintrans_cosmo} only differ by a few percent if applied to the control simulation. Therefore, equation~\ref{equ:spintrans_cosmo} can be regarded as a consistent extension of equation~\ref{equ:deltalambda_ctl} to cosmological simulations.

\subsection{Comparing control and cosmological runs}

\begin{figure*}
    \centering
    \includegraphics[width=\textwidth]{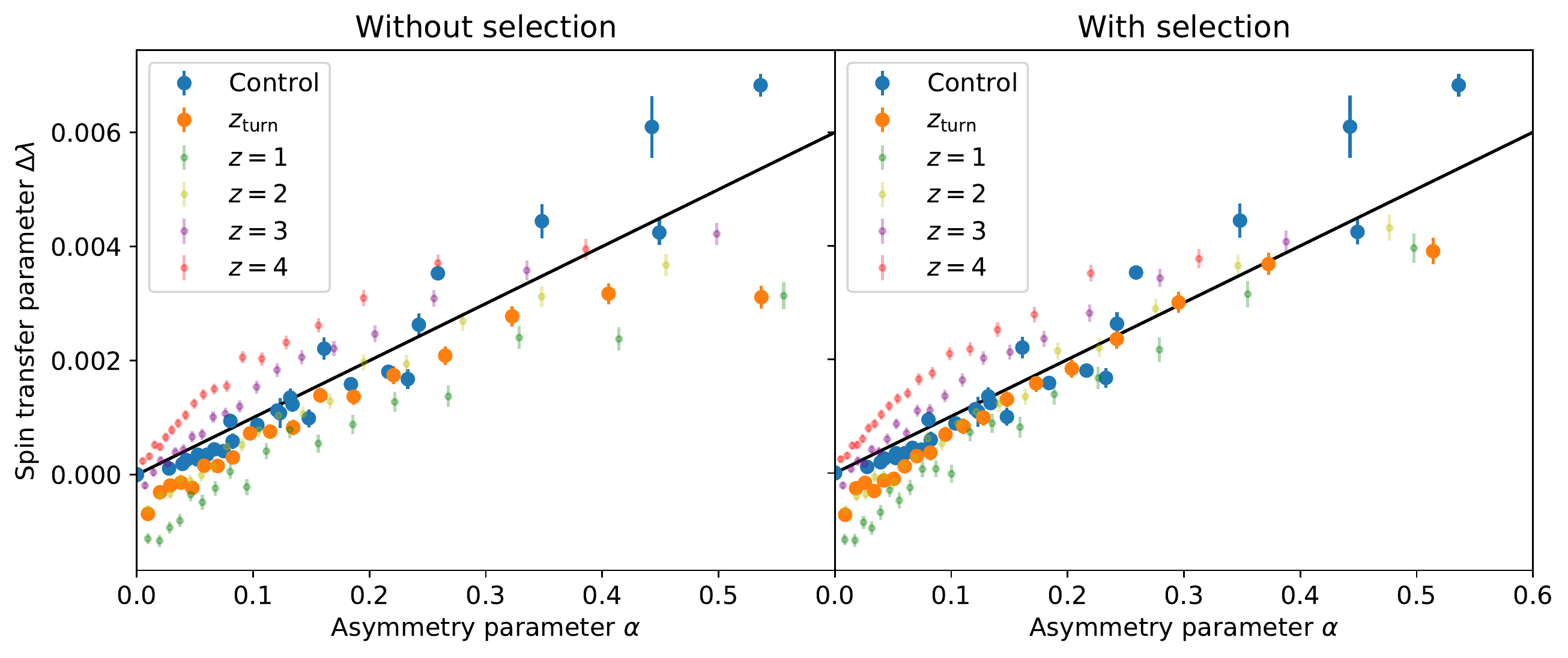}
    \caption{$\alpha-\Delta \lambda$ relation of control (blue points) and cosmological simulations (other colored points). The orange points refer to the SUFRS haloes at $z_{\rm turn}$, when the physical rms radius is at its maximum. Other points are for fix redshifts $z=1,2,3,4$. The straight line is the fit $\Delta \lambda = 0.01\alpha$. The errorbars show the uncertainty of the mean value at each $\alpha$ bin. {\it Left panel:} All SURFS haloes with $M_{\rm halo}\geqslant 10^{12}h^{-1}\Msun$. {\it Right panel:} Only SURFS haloes with initial axies ratio $\avg{x_i}/\avg{y_i}\leq 2$.}
    \label{fig:AsymPara}
\end{figure*}

Figure~\ref{fig:AsymPara} presents the average $\alpha-\Delta \lambda$ relation in SURFS, in $\alpha$-bins of identical numbers of haloes. The control simulations are also shown in this figure. We focus on the average relation in SURFS, because our model cannot make a strong prediction on a halo-by-halo basis due to the large variety in geometry and assembly histories in the cosmological simulation. These factors cause a large scatter of the spin transfer parameter $\Delta \lambda$ of about 0.7 dex. 

The left panel of Figure~\ref{fig:AsymPara} shows that the average spin transfers $\Delta \lambda$ between $z_{\rm turn}$ and $z=0$ in SURFS (orange points) lie close to those computed from the ellipsoidal top-hat collapse (blue points), given identical initial asymmetry parameters $\alpha$. This is rather remarkable given the simplicity of the ellipsoidal model compared to real halos. The right panel shows the same comparison, but only for SURFS haloes whose Lagrangian region at turn-around has a major-to-minor axes ratio below 2 (in the plane orthogonal to the spin). Larger axes ratios are often 
associated with Lagrangian regions that undergo major mergers before forming a single halo. Such systems are not expected to be well-described by a single ellipsoid. The figure demonstrates that, if such cases are ignored, the agreement between the cosmological simulation and the ellipsoidal control runs is further improved. These results are a strong indication that the ellipsoidal collapse model captures the essence of the systematic transfer of AM from DM to gas in non-radiative cosmological simulations.

For comparison, Figure~\ref{fig:AsymPara} also shows the $\Delta \lambda$ values between other redshifts ($z=1,2,3,4$) and $z=0$ (light-coloured points). These values are not expected to be well-matched by the ellipsoidal model, which starts at the turn-around. The points associated with $z=2$ show the best agreement with the control runs and hence with the linear model (black line), because the mean turn-around redshift of our sample lies close to $z=2$. At the lower redshifts, some haloes with small $\alpha$ have already collapsed. Substructures within the haloes then dominate the AM transfer at this stage, which may cause the gas to transfer excess AM back to the DM. This explains why $\Delta \lambda<0$ at small $\alpha$, especially at low $z$. In turn, at higher redshifts ($z>2$), most Lagrangian regions associated with $z=0$ haloes are still expanding and have not yet reached their turn-around point. This decreases their measured collapse factor and thus underestimates their $\alpha$. This causes these haloes to land left (or on top) of the reference $\alpha-\Delta \lambda$ relation.

\begin{figure}
    \centering
    \includegraphics[width=\columnwidth]{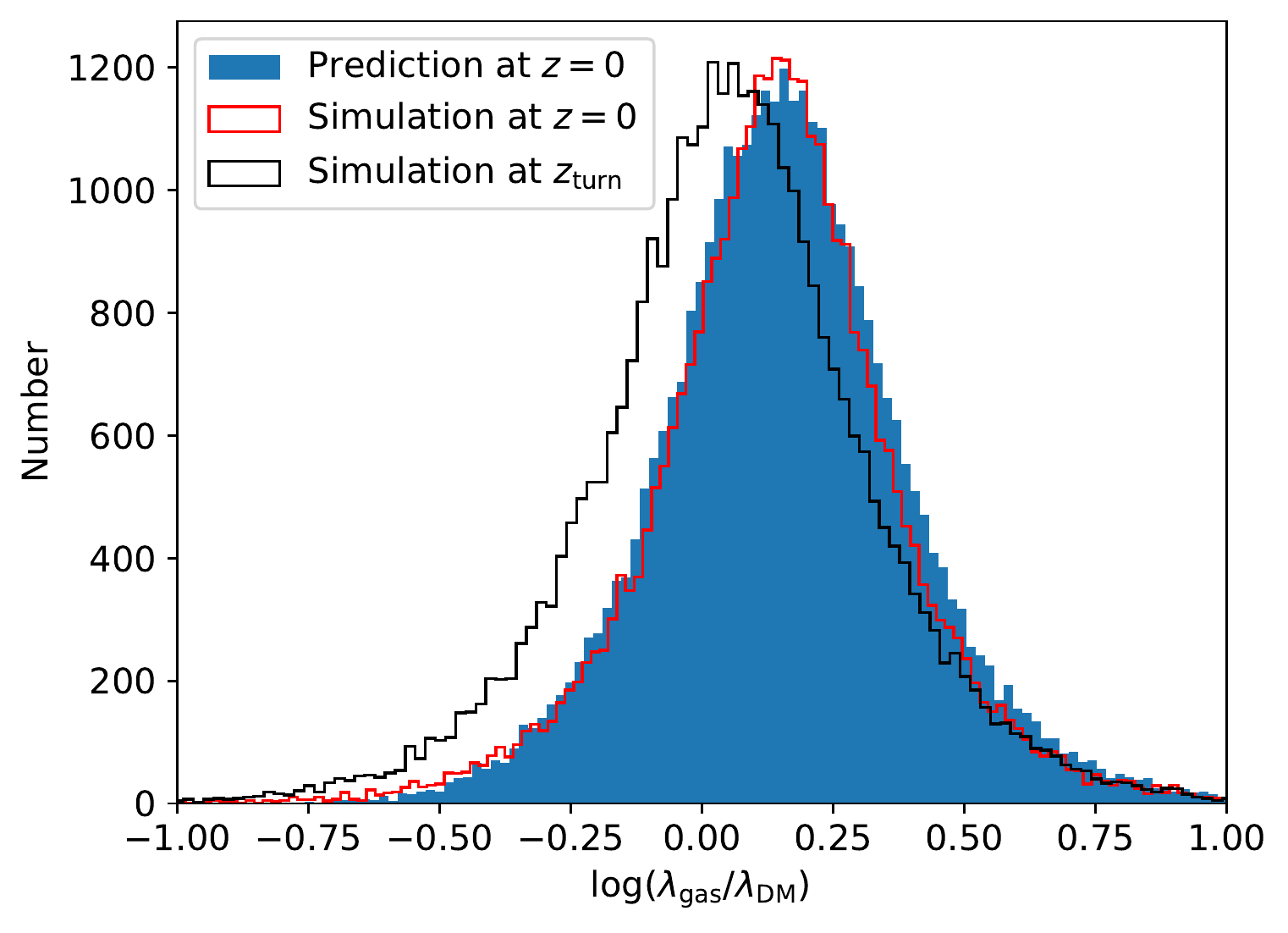}
    \caption{Blue shows the distribution of $\xspingas/\xspinDM$ predicted by equation~\ref{equ:predictj} from SURFS. This distribution is very similar to the true distribution (red line).}
    \label{fig:predictlambda}
\end{figure}

 Given this success, can we then use this model to explain the value of $\avg{\xspingas}/\avg{\xspinDM}>1$ at $z=0$ in Section~\ref{s:cosmosims}? To answer this question, we use this model to predict the final AM ratio at $z=0$, for each halo, based on their initial spins. Explicitly, we evaluate the equation
\begin{equation}
    \frac{\xspingas(t_f)}{\xspinDM(t_f)}=\frac{\jg(t_f)}{\jd(t_f)}=\frac{\jg(t_i)+\f^{-1}\Delta j}{\jd(t_i)-(1-\f)^{-1}\Delta j},
    \label{equ:predictj}
\end{equation}
which follows from equation~\ref{equ:ctl_totalAMtransfer4} assuming $\Delta J_{\rm{tot,ext}}=0$. In line with the assumptions of equation \ref{equ:ctl_totalAMtransfer2}, we here made the approximation that the AM of gas and DM are parallel. Previous studies (\citealp[e.g.][]{van_den_bosch_angular_2002,sharma_origin_2012,bett_angular_2010}) found that the angle between these vectors lies around $20^{\circ}$--$30^{\circ}$, on average, meaning that we make a $\sim15\%$ error if considering only vector norms. This is an acceptable systematic error in view of the statistical scatter of about 0.7 dex in $\xspingas/\xspinDM$ (see Figure \ref{fig:predictlambda}).

The value of $\Delta j$ in equation \ref{equ:predictj} is then substituted for $\Delta \lambda$ using equation~\ref{equ:spintrans_cosmo}, which can be computed from computed from $\alpha$ (as computed from the spin and geometry at $z_{\rm turn}$) via equation~\ref{equ:deltalambdaalpha}.

The result of this model prediction is shown in Figure~\ref{fig:predictlambda}. The predicted distribution at $z=0$ (blue histogram) is indeed very similar to the true distribution at $z=0$, shown as the red line. Also shown in this figure is the distribution at $z_{\rm turn}$. The large width of this distribution, though quite symmetrical in the exchange of DM and gas, hints at the complex assembly histories and torques experienced by different haloes. It is this complex variety that limits the use of our model to average relations and distributions. We briefly discuss the effect of the assembly histories in Section~\ref{ss:linkfullphysics}.

\subsection{Beyond adiabatic physics}
\label{ss:linkfullphysics}

The AM of the gas in real haloes is expected to differ from the predictions of non-radiative physics for two main reasons.

First, complex baryonic physics, especially cooling and feedback from stars and supernovae, can have significant effects on the AM of the gas. These effects have been studied quite extensively, at the level of galactic discs \citep[e.g.][]{stevens_how_2017,stewart_angular_2013,governato_bulgeless_2010}, but less so at the level of the halo gas. It is, however, natural to assume that the predictions from non-radiative physics (as in this paper) would be mainly applicable to very massive haloes ($M\gtrsim10^{13}\Msun$), in which the vast majority of the baryonic component exists as a hot atmosphere, not significantly affected by cooling.

As the cooling may decrease the AM of gas, feedback enhances it. \citet{zjupa_angular_2017} suggested that in the simulations with active galaxy formation physics, the baryonic spin increases to an average ratio $\xspingas/\xspinDM \sim 1.8$. By assuming that the low AM gas is removed by galactic winds and AGN feedback, they estimated the ratio $\xspingas/\xspinDM \sim 1.55$ from these effects, which indicates that feedback is a significant factor in increasing the gas spin.

Second, environmental effects, such as the effects of filaments and hierarchical assembly, of real haloes are likely to play important roles in setting the spin and gas-to-DM spin ratio of haloes. Also, major mergers are expected to bring in a lot of orbital angular momentum, while also dissociating the gas from the DM via ram pressure, hence causing torques between these components. We will focus on the role of this assembly history on the gas-to-DM spin ratio in an upcoming paper.

\section{Conclusions}
\label{s:conclusions}

We have studied the spin of DM and gas in haloes formed in cosmological simulations in the absence of cooling and galaxy formation. Even in this simplified setting, the sAM of the gas is, on average, significantly higher than that of the DM, with $\avg{\jg}/\avg{\jd}\approx1.44$ in haloes at $z=0$. We found that most of this excess AM lies in the inner halo parts ($<0.5R_{\rm vir}$) and arises after the turn-around point during halo formation. We also found that the evolution of $\avg{\jg}/\avg{\jd}$ is associated with a cosmic evolution in the gas spin distribution, whereas the spin of the DM component remains nearly constant with time.

Through a series of control simulations of an ellipsoidal top-hat
collapse, we have identified the leading mechanism behind the origin of the excess spin of the gas. As the asymmetric halo collapses, the
pressurised inner gas shells implode more slowly, causing the DM to spin slightly ahead of the gas. This thus induced torque systematically transfers AM from the DM to the gas. This model predicts that the amount of AM transferred to the gas, as quantified by the dimensionless `spin transfer parameter' $\Delta \lambda$, scales approximately linearly with a dimensionless `asymmetry parameter' $\alpha$ that only depends on the initial axes ratios, spin parameter and collapse factor. Remarkably, the $\alpha$--$\Delta\lambda$ relation fitted to the ellipsoidal top-hat
model (equation~\ref{equ:deltalambdaalpha}), closely matches the average $\alpha$--$\Delta\lambda$ relation seen in a non-radiative cosmological simulation. This corroborates the claim that, in the absence of cooling physics, most of the average excess of AM in the gas is explained by this model.

The gas-to-DM spin ratio of haloes is likely to affect the formation and evolution of galaxies, as the halo gas seeds the star-forming gas in galactic discs. Thus, semi-analytic models would be well-advised to adopt a slightly higher spin for the pristine halo gas (e.g. following equation~\ref{equ:jratio_z_fit}) than what would be inferred from DM-only simulations. However, we note that in the presence of galaxies the story becomes more complex, since the halo gas is intimately connected to galaxies via complex feedback mechanisms.

Finally, it is worth repeating that our model is not particularly
suitable to explain the gas-to-DM spin ratio of individual haloes,
because local environmental differences and different assembly histories cause large halo-to-halo variations, even amongst haloes with the same value of $\alpha$. Investigating how different assembly histories affect the gas-to-DM spin ratio on a halo-to-halo basis is therefore an interesting avenue to pursue further.



\section*{Data Availability}

The data used in this article were generated using the National Computing Infrastructure (NCI) high performance computing facility in Canberra, Australia. The derived data generated in this research will be available upon request.

\section*{Acknowledgements}
DO is a recipient of an Australian Research Council Future Fellowship (FT190100083) funded by the Australian Government. CL has received funding from the ARC Centre of Excellence for All Sky Astrophysics in 3 Dimensions (ASTRO 3D), through project number CE170100013, and an ARC Discovery Project, through project number DP210101945. CL also thanks the MERAC Foundation for a Postdoctoral Research Award. CP acknowledges the support of ASTRO 3D. We thank the anonymous referee for very constructive comments.




\bibliographystyle{mnras}




\appendix

\section{Resolution for the control simulations}\label{app:convergence}

We assess the number of required simulation particles through a convergence analysis of the particular ellipsoidal top-hat collapse shown in Figure~\ref{fig:CtrlSim_one_example}. Its initial conditions are characterised by the axes ratios $a_x:a_y:a_z=1.8:1.0:1.0$, a dispersion parameter $\kappa=0.2$ and spin parameter $\lambda=0.06$. We chose this halo for the convergence study, because it produced the highest torques between gas and DM of all considered control runs and is therefore most sensitive to subsampling inaccuracies.

\begin{figure}
    \centering
    \includegraphics[width=\columnwidth]{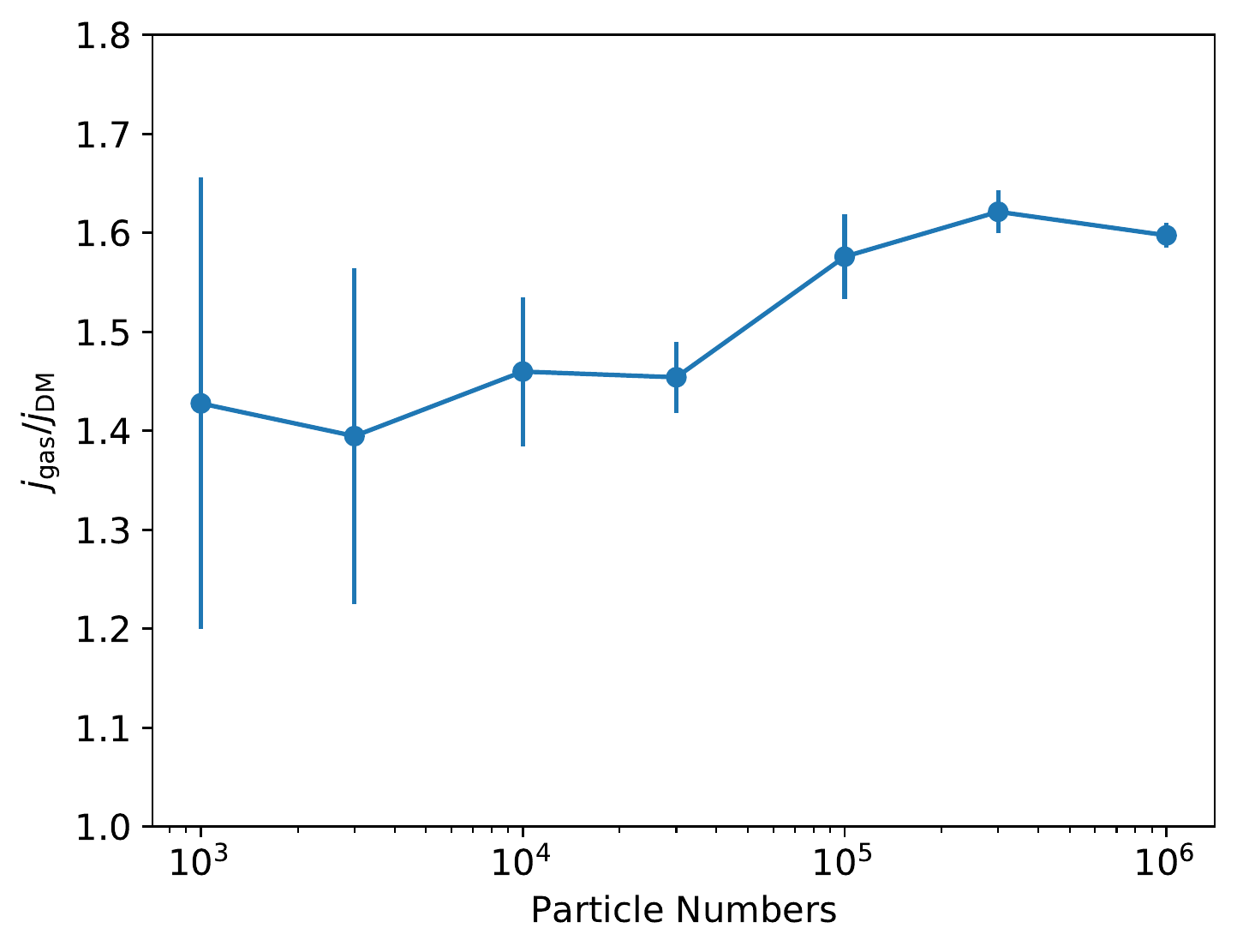}
    \caption{Final \jgas/\jDM ratio of a the particular ellispoidal collapse shown in Figure~\ref{fig:CtrlSim_one_example}, as a function of the total number of particles used for the simulation. For each particle number, we considered five random seeds, giving rise to a spread in the spin ratio shown by the 1-$\sigma$ error bars.}
    \label{fig:resolution}
\end{figure}

For the convergence analysis, we have resimulated this halo while varying the total number of particles $n_{\rm p}$ from $10^3$ to $10^6$. The results are shown in Table~\ref{tab:resolution} and Figure~\ref{fig:resolution}. The error bars on the final \jgas/\jDM ratios are standard deviations of five random seeds for each number of particles.

This analysis reveals that as few as $10^3$ particles consistently predict a transfer of AM from the DM to the gas. However, for the smaller numbers of particles, the final spin ratio \jgas/\jDM is somewhat biased towards lower values. At $10^5$ particles, this bias has disappeared and the statistical scatter is small enough for the purposes of this paper. We keep track of this statistical scatter by simulating each control run with five seeds throughout this paper.

\begin{table}
    \centering
    \begin{tabular}{cccc}
       \hline
       $n_{\rm p}$  & $m_{\rm p}$ & \jgas/\jDM & $\Delta(\jg/\jd)$  \\
       \hline
       $1\times 10^3$ & $1\times 10^9$ & 1.43 & 0.23  \\
       $3\times 10^3$ & $3.33\times 10^8$ & 1.39 & 0.17  \\
       $1\times 10^4$ & $1\times 10^8$ & 1.46 & 0.08  \\
       $3\times 10^4$ & $3.33\times 10^7$ & 1.45 & 0.04   \\
       $1\times 10^5$ & $1\times 10^7$ & 1.58 & 0.04  \\
       $3\times 10^5$ & $3.33\times 10^6$ & 1.62 & 0.02  \\
       $1\times 10^6$ & $1\times 10^6$ & 1.59 & 0.01  \\
       \hline 
    \end{tabular}
    \caption{Column 1: the number of particles; Column 2: the particle mass; Column 3: the ratio of \jgas and \jDM; Column 4: the standard deviation of 5 random seeds of each sample.}
    \label{tab:resolution}
\end{table}

\bsp	
\label{lastpage}
\end{document}